\documentclass{jpp}
\usepackage{graphicx}
\usepackage[T1]{fontenc}
\usepackage{amsmath}

\usepackage{booktabs}
\usepackage{multirow}
\usepackage{threeparttable}

\newcommand{\myrefeq}[1]{Eq.~(\ref{#1})}
\newcommand{\myreftab}[1]{Table~\ref{#1}}
\newcommand{\myreffig}[1]{Fig.~\ref{#1}} 

\shorttitle{Radio wave emission by electrn beam plasmas}
\shortauthor{Xin Yao, {Patricio A. Mu\~{n}oz}, {J\"{o}rg B\"{u}chner}, {Xiaowei Zhou} and {Siming Liu}}

\title{The effects of density inhomogeneities on the radio wave emission in electron beam plasmas}

\author{Xin Yao\aff{1,2}\corresp{\email{xin.yao@campus.tu-berlin.de}},
        {Patricio A. Mu\~{n}oz}\aff{2},
        {J\"{o}rg B\"{u}chner}\aff{2,1},
        {Xiaowei Zhou}\aff{3,1},
        \and {Siming Liu}\aff{3}}

\affiliation{\aff{1}Max Planck Institute for Solar System Research, 37077 {G\"{o}ttingen}, Germany
\aff{2}Centre for Astronomy and Astrophysics, Technical University of Berlin, 10623 Berlin, Germany
\aff{3}Key Laboratory of Dark Matter and Space Astronomy, Purple Mountain Observatory, Chinese Academy of Sciences, 210034 Nanjing, China}

\begin{document}

\maketitle


\begin{abstract}
\normalsize
Type III radio bursts are radio emissions associated with solar flares. They are considered to be caused by electron beams traveling from the solar corona to the solar wind. Magnetic reconnection is a possible accelerator of electron beams in the course of solar flares since it causes unstable distribution functions and density inhomogeneities (cavities). The properties of radio emission by electron beams in an inhomogeneous environment are still poorly understood. We capture the non-linear kinetic plasma processes of generation of beam-related radio emissions in  inhomogeneous plasmas by utilizing fully-kinetic Particle-In-Cell (PIC) code numerical simulations. Our model takes into account initial electron velocity distribution functions (EVDFs) as they are supposed to be created by magnetic reconnection. We focus our analysis on low-density regions with strong magnetic fields. The assumed EVDFs allow two distinct mechanisms of radio wave emissions: plasma emissions due to wave-wave interactions and so-called electron cyclotron maser emissions (ECME) due to direct wave-particle interactions. We investigate the effects of density inhomogeneities on the conversion of free energy from the electron beams into the energy of electrostatic and electromagnetic waves via plasma emission and ECME, as well as the frequency shift of electron resonances caused by perpendicular gradients in the beam EVDFs. Our most important finding is that the number of harmonics of Langmuir waves increases due to the presence of density inhomogeneities. The additional harmonics of Langmuir waves are generated by a coalescence of beam-generated Langmuir waves and their harmonics. 
\end{abstract}

\section{Introduction}\label{sec:introduction}

Solar flares are the most energetic phenomena observed in the Sun. They are thought to be powered by magnetic reconnection, in the course of which magnetic energy is converted to other forms of energy \citep{Treumann2013}, in particular to accelerated electrons propagating as beams. These electron beams can generate electromagnetic waves in a wide range of wavelengths. For example, while precipitating to the solar chromosphere, energetic electrons can cause X-ray emissions. On the other hand, outward escaping electrons emit radio waves along their way~\citep[][]{Reid2014, Melrose2017a}. 
Among electromagnetic waves non-thermal, coherent radio emissions are very characteristic. Their brightness temperature can exceed the equivalent thermal (blackbody) radiation temperature by orders of magnitude. In contrast to incoherent radiation, non-thermal electromagnetic waves are generated by collective plasma processes. Since corresponding solar radio emissions are inherently linked to plasma processes, in the regions where solar flares take place they can be used as remote diagnostic tools to better understand the physics of solar eruptions and the magnetic reconnection processes behind. 
In particular solar flares are associated with the so-called type III solar radio bursts (SRBs), which are believed to be caused by electron beam instabilities~\citep[see, e.g.,][and references therein]{Reid2014}. The source of free energy for such emissions can be found in the electron velocity distribution functions (EVDFs). In order to understand the solar radio emissions, it is therefore of critical importance to analyze the wave emission properties of electron beams due to EVDFs caused by magnetic reconnection.

Type III SRBs are characterized by their spectral structure of fundamental and harmonic wave modes. Radio observations reveal that solar flares emit radio waves not only at the local plasma frequency $\omega_{pe}$ at the source region, but also at their first and even higher harmonic wave modes  $n\omega_{pe}, n=2,3,\dots$~\citep[][]{Smerd1976,Reiner1992}. This implies that they could be caused by wave-wave interactions~\citep[see, e.g.,][and references therein]{Reid2014, Melrose2017a}. 
Indeed, the most widely accepted mechanism of SRBs is the \emph{plasma emission} mechanism, which relies on wave-wave coupling of plasma waves induced by electron beam instabilities~\citep[e.g., see ][]{Ginzburg1958,Melrose1970,Melrose1970a}. The electron beams firstly generate Langmuir waves via a bump-on-tail instability due to a positive gradient in the EVDF parallel to the magnetic field direction: $v_{\parallel}\cdot\partial f/\partial v_{\parallel}>0$. 
The resulting Langmuir ($L$) wave intensities usually exceed the level of thermal excitations by orders of magnitude. In the course of a multistage process these beam-generated Langmuir waves can interact with ion-sound ($S$) waves, another normal wave mode of the plasma, and electromagnetic waves ($T$) could be generated by their interaction according to, e.g, ~\citet[][]{Ginzburg1958,Melrose1970,Melrose1970a}. These electromagnetic waves can finally escape the plasma and be observed remotely, provided their frequency is higher than the (cutoff) local plasma frequency.

Waves generated by plasma emission mechanism do not depend on the strength of the magnetic field at their source region. However, some features of type III SRBs, e.g., O mode polarized waves, depend on the strength of the coronal magnetic field in the source region of the emission. In fact, magnetic fields allow wave generation due to the interaction of waves with the cyclotron motion of particles, in particular of electrons. Based on a wave-particle interaction, an alternative cyclotron-resonance related mechanism named \emph{electron cyclotron maser emission} (ECME) was proposed ~\citep[see, e.g.,][and references therein]{Twiss1958,Wu1979,Treumann2006}. The cyclotron resonance condition is given by $\omega-k v_{\parallel}=n\Omega_{ce}/\gamma$, where $\Omega_{ce}$ is the electron cyclotron frequency and $\gamma$ is the relativistic Lorentz factor. Such resonances are possible in the solar corona despite the coronal conditions in which the electrons are only mildly relativistic. For ECME a positive gradient in the EVDF is needed in the direction perpendicular to the magnetic field: $\partial f/\partial v_{\perp}>0$, e.g., in loss-cone~\citep{Benacek2017}, ring~\citep{Pritchett1984,Lee2011}, horseshoe~\citep{Bingham2000,Melrose2016b}, cup-like~\citep{Buchner1996} or shell-shaped distribution functions. The corresponding ``inverted'' population in the velocity space led to the early authors call this cyclotron-resonance related mechanism a ``maser'' mechanism. The characteristic frequencies of ECME are generally at $\Omega_{ce}$ and at their harmonics $n\Omega_{ce}$ ($n>1$). This mechanism also requires the condition $\Omega_{ce}>\omega_{pe}$ for the resulting waves to escape from the plasma. This frequency condition is actually the opposite of the one typical in the solar corona. It implies a sufficiently low plasma density ($\omega_{pe}\propto \sqrt{n_e}$, with $n_e$ the electron plasma density) in regions of strong magnetic fields ($\Omega_{ce}\propto B$, with $B$ the background magnetic field strength). However, this condition can be met in density cavities (i.e. regions of locally depleted plasma) in the solar corona, e.g., in regions associated with hard-X ray bursts during solar flares~\citep{Regnier2015,Melrose2016b}. Observations by \citet{Regnier2015,Morosan2016} have confirmed the condition $\Omega_{ce}>\omega_{pe}$ for ECME in the solar corona by a method combining magnetic field extrapolation techniques and hydrodynamic models. In addition, numerical simulations have demonstrated that density cavities form along one pair of the separatrices of guide-field magnetic reconnection~\citep{Ricci2004,Pritchett2004,Munoz2016}. 
It has recently been theoretically proposed that the electron cyclotron maser instability (ECMI) can be operative in those separatrices regions of magnetic reconnection \citep{Treumann2017}.

In order to understand the different aspects of plasma emission and ECME, a number of studies have already been carried out~\citep[see, e.g.,][and references therein]{Melrose2017a,Zhou2020}. 
Those studies usually focused on either EVDFs with positive gradients in the parallel direction (for plasma emission) or in the perpendicular direction (for ECME emission) to the magnetic field. 
More complicated EVDFs providing both kinds of free energy are, however, expected to be formed by magnetic reconnection.
So we consider a simple distribution function model, namely a ring-beam distribution, that can provide free energy for both instabilities.
Note that an important observational distinction between both instabilities is that the plasma emission favours O mode emission~\citep[][]{Melrose2017a,Melrose1978}, while the electron cyclotron emission strongly prefers the X mode ~\citep[][]{Ellis1962,Melrose2017a}.

Ring-beam distribution functions can be formed through a variety of mechanisms. They can be generated due to magnetic gradient drifts, e.g. when plasma jets (beams) cross a tangential discontinuity in the magnetic field~\citep{Voitcu2018}. Such scenario applies when steep magnetic gradients are formed as in collisionless shocks and by magnetic reconnection at kinetic scales. Magnetic gradient drifts can redistribute the parallel beam particle energy into the perpendicular direction~\citep{Zhou2015a}, thus forming ring-beam and gyro-phase restricted EVDFs~\citep{Voitcu2012} as well as crescent-shaped EVDFs~\citep{Voitcu2018}. Ring distributions have been indeed found in Particle-in-cell (PIC) simulations of magnetic reconnection~\citep{Shuster2014,Bessho2014} and in quasi-perpendicular shocks~\citep{Tokar1986}. In the latter case they were formed via surfatron and shock acceleration~\citep{Bingham2003}.  Magnetic reconnection in strong magnetic fields has also been proposed to be able to generate perpendicular gradients in the EVDFs along the separatrix regions~\citep{Treumann2017}. 
A linear stability analysis of ring-beam distributions was carried out, e.g., by~\citep{Vandas2015}. 
The non-linear evolution of the wave excitation due to ring-beam distribution functions was investigated in detail, e.g., by~\citep{Zhou2020}. Those authors derived the properties of waves generated by ring-beam distribution functions in the solar coronal plasma, like the polarization of escaping waves.

In solar flares, related electron beams formed by magnetic reconnection were reported to take place in regions of strong turbulence and corresponding density fluctuations both at large scales (e.g., compressible magnetosonic MHD wave turbulence) and (sub-ion) kinetic scales~\citep{Drake2003,Munoz2018b}. This turbulence and its associated density fluctuations will influence the radio waves by modifying, e.g., the local plasma frequency and thus the escape condition for electromagnetic waves, the wave-particle resonance condition of cyclotron resonances and Landau damping, possibly even enhancing the unstable growth of plasma waves~\citep{Wu2012}, and providing additional channels for wave generation via mode conversion process~\citep{Kim2007}. 
Mode conversion from Langmuir waves to escaping electromagnetic waves due to random density irregularities is particularly important as it was found by quasi-linear and test particle investigations~\citep{Cairns2005,Krafft2015,Volokitin2018,Krasnoselskikh2019,Krafft2020}. 
Another proposed conversion process from electrostatic to electromagnetic waves is the antenna emission process, according to which Langmuir waves can become trapped in localized density wells, where they can be converted to electromagnetic radiation~\citep{Malaspina2012}. 
Radio emission due to electron beams by magnetic reconnection was simulated utilizing PIC codes~\citep{Sakai2005}. PIC code simulations of localized beams for laboratory experiment conditions demonstrated that electromagnetic waves can be generated very efficiently at the first harmonic of the plasma frequency by an antenna mechanism due to wave-coupling with ion-acoustic waves~\citep{Annenkov2019a}. 

There is a second consequence expected to be due to solar flare-related magnetic reconnection --- the formation of density inhomogeneities~\citep{Ricci2004,Pritchett2004,Munoz2016}.
In order to simplify both theoretical and numerical calculations, previous investigations of plasma emission and ECME were carried out for a homogeneous coronal plasma background in the source regions. Despite the fact that density inhomogeneities are perhaps a very common situation in the source region of solar flares, it has not yet been clarified by a first-principle kinetic approach how density gradients affect the resulting radio emission. 
Previous studies of density gradients effects on electron radio emissions focused on large scale inhomogeneities. 
For example, \citet{Tsiklauri2011b,Pechhacker2012c,Schmitz2013b} studied, via 1D PIC simulations, wave emission by electron beams with density gradient scales inspired by inhomogeneities measured at solar wind length scales. 
However, the influence of small-scale density gradients on the properties of radio emission due to plasma emission and ECME mechanisms by electron beams is still poorly understood.
The effects of random density inhomogeneities on wave emission have just recently started to be investigated by a fully-kinetic model for solar wind conditions~\citep{Thurgood2016}.

In order to find out how small-scale densities inhomogeneities affect beam-related radio emissions generated in the solar corona, we have now extended the analysis of beam instabilities causing radio emissions to the consideration of the consequences of density gradients at sub-ion (kinetic) scales. 
Further we note that previous studies of electron beam related radio emissions focused on the parameter regime $\Omega_{ce}<\omega_{pe}$, where electron cyclotron waves cannot efficiently escape \citep[e.g.,][]{Tsiklauri2011b,Pechhacker2012c,Schmitz2013b}. In our study, we instead focus on the opposite parameter regime $\Omega_{ce}>\omega_{pe}$, like in density cavities generated by magnetic reconnection, in which escaping waves can be directly generated.

The standard plasma emission mechanism (three-wave interaction mediated by ion sound waves) predicts the generation of waves at the fundamental plasma frequency. Proposed already more than half a century ago, it has recently been unambiguously confirmed by kinetic numerical simulations \citep{Thurgood2015,Henri2019}.
However, higher order harmonic waves were also observed in SRBs~\citep[][]{Takakura1974,Reiner2019a}. Various theories were developed to explain them~\citep[]{Gaelzer2002,Yoon2003a,Yoon2005,Yi2007,Rhee2009}. 
While harmonics of Langmuir waves were claimed to be found in some numerical experiments \citep[e.g.,][]{Rhee2009}, other numerical studies of wave generation failed to find conclusive evidence for the generation of harmonics due to the standard plasma emission mechanism~\citep{Ganse2012,Zhou2020}. 
Fully-kinetic PIC simulations of plasma emission leading to the generation of higher order harmonic(s) of the plasma frequency waves are very rare. This is attributed to the numerical difficulties and their sensitive dependence to plasma parameters like the beam density. 

By our fully-kinetic PIC simulations, we now have found that waves at the harmonics of the local plasma frequency ($n\omega_{pe},n=1,2,\dots$) can be generated by electron beams and their properties depend on the background density gradients.
Those harmonics are due to the non-linear interaction of beam-generated Langmuir waves with adjacent lower order harmonics, i.e., $L+L_n\to L_{n+1}$.

Note that our study focuses on a local generation mechanism of Type III solar radio bursts due to magnetic reconnection generating superthermal electrons in solar flares. 
While some aspects of this process have already been investigated previously~\citep{Wu2014,Treumann2017,Zhou2020}, we analyzed here the generation mechanisms of fundamental of Langmuir and electron cyclotron waves as well as their harmonics when the electron beams propagate through small-scale density gradients in the ambient plasma. We show that density inhomogeneities fundamentally change the properties of the resulting waves. 
For a direct comparison with Type III SRBs observations, the mechanism studied here will have to be combined with a global model to take also into account the wave propagation and electron transport effects in the solar corona and in the solar wind \citep[see, e.g,][]{Li2009,Reid2018}.
Because of the short-time scales that we investigated, our results are best suited for short duration and high-frequency Type III SRBs, like radio spikes \citep[see, e.g.,][and references therein]{Fleishman1998}.

The paper is organized as follows: our numerical model and simulation setup as well as the chosen parameters are presented in Section~\ref{sec:model}, the results of our simulations about fundamental and harmonics of Langmuir and electron cyclotron waves are presented in the Section~\ref{sec:results}, and our conclusions are summarized in Section~\ref{sec:conclusions}. Convergence tests are discussed in the Appendix~\ref{app:CT}.

\section{Numerical Model}\label{sec:model}

Our investigations utilized the fully-kinetic Particle-in-Cell (PIC) code ACRONYM~\citep[][]{Kilian2012} in its 2.5-dimensional version. This means a two-dimensional mesh grid in space is used while the full three-dimensional particle motion is taken into account. The code numerically solves the Vlasov equation and thus it is appropriate to model the collisionless plasmas of the solar corona.

\begin{figure}
    \centering
    \includegraphics[width=0.85\textwidth]{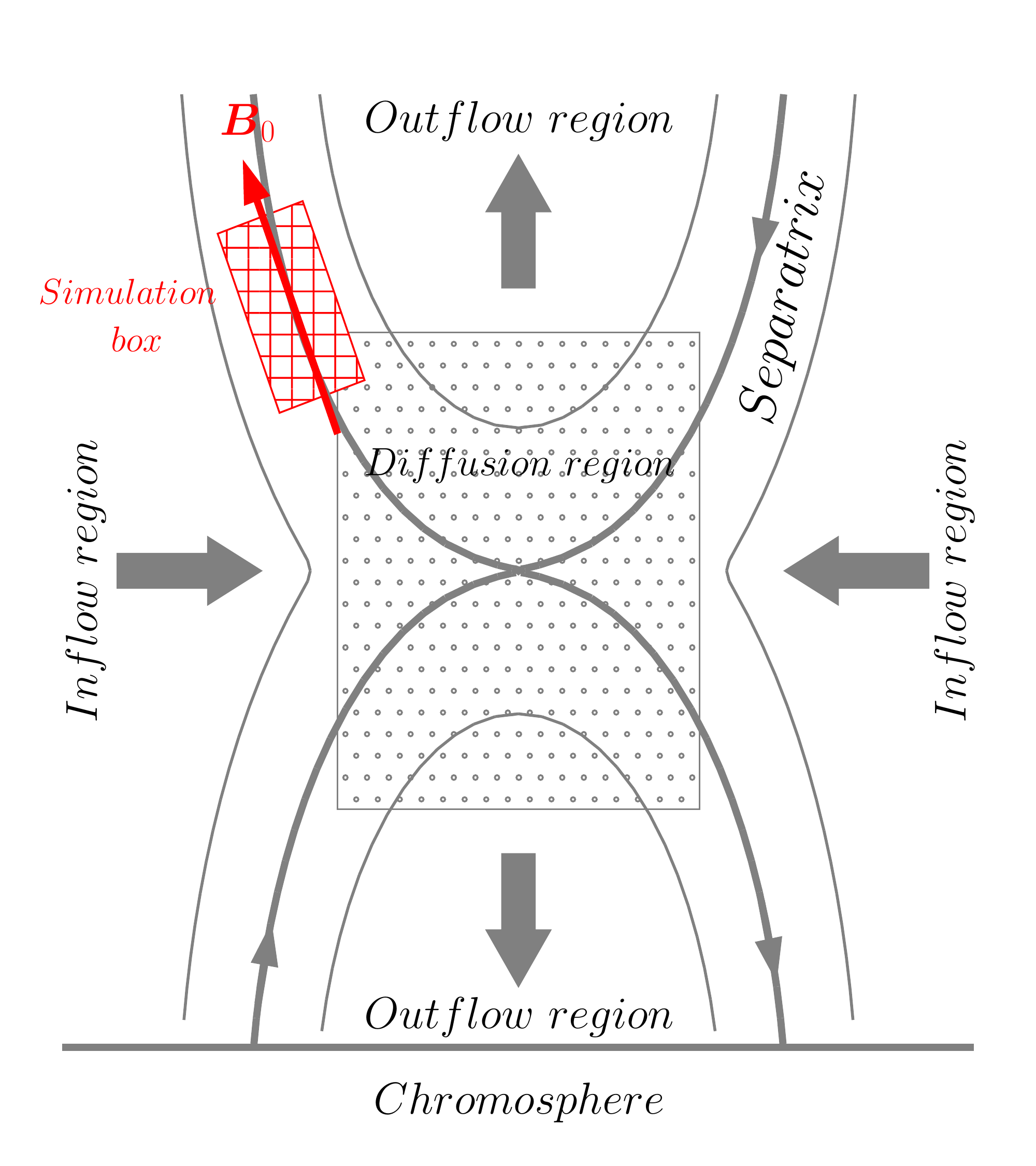}
    \caption[]{Schematic of the simulation box (red rectangle) in the separatrix region of  magnetic reconnection in solar flares. The box extends along the direction of the local solar magnetic field $\boldsymbol{B}_0$.}
    \label{fig:simulationbox}
\end{figure}

We studied the radio emission caused by electron beams as they might have been accelerated in solar flares in a 2D simulation box located in the $x\ vs.\ y$ plane (see \myreffig{fig:simulationbox}). This simulation domain represents part of the separatrix region of magnetic reconnection, and it extends along the direction of the local solar magnetic field $\boldsymbol{B}_0$. The background magnetic field is assumed to be constant throughout the box, and the direction of the magnetic field defines the $x$ direction of the simulation box, i.e., $\boldsymbol{B_0}=B_0\boldsymbol{e}_x$ (see \myreffig{fig:simulationbox}). From here on we refer to the $x$ direction as the parallel direction. 

We considered an electron-proton plasma with a realistic proton-to-electron mass ratio $m_p/m_e=1836$.
The plasma consists of a background and an electron beam population streaming at a given drift speed (which is described in more detail below).  The initial electron plasma frequency is set to be $\omega_{pe}=5.0\times 10^9\ rad/s$, which corresponds to an electron number density of $n_0=7.9\times 10^9\ cm^{-3}$,
typical for the solar corona~\citep[][]{Aschwanden2005}. The ratio of the electron cyclotron frequency $\Omega_{ce}$ to the electron plasma frequency $\omega_{pe}$ is $\Omega_{ce}/\omega_{pe}=4$, corresponding to a magnetic field $B_0=1137 G$. In such an environment the ECME instability could generate radio waves. Even though this frequency ratio is not common for the solar coronal plasma,
it applies to cavity ducts in which the density is lower than that in the  surrounding coronal plasma~\citep[][]{Morosan2016,Zhou2020}. The thermal speed of the background electron plasma is $v_{the}=0.025\ c$, which corresponds to a coronal plasma temperature $T_{bg}=3.7\times10^{6}\ K$ and results in a Debye length of $\lambda_D=0.15\ cm$.

The size of the simulation box is $(L_x,L_y)=(N_x,N_y)\times\Delta x$ along the $x$ and $y$ directions, respectively. We set the grid cell size to be $\Delta x=2\lambda_D$ and the number of grid points is $N_x=4096, N_y=512$. Periodic boundary conditions are applied in both directions of the simulation box.
In order to satisfy the Courant-Friedrichs-Lewy (CFL) condition, we imposed  the condition $c\Delta t/\Delta x=1/2<1/\sqrt{3}$, thus $\Delta t=0.025\ \omega_{pe}^{-1}$. 
Based on a sampling period of $5\Delta t$, our simulations allow to obtain frequencies up to $\omega=24\ \omega_{pe}$ or  $\omega=6\ \Omega_{ce}$ in the frequency domain. The dispersion relation analysis (to be presented later) are always based on a time window which allows a frequency resolution of $\Delta \omega=0.047\ \omega_{pe}$ or $\Delta \omega=0.012\ \Omega_{ce}$. 
The size of the simulation box allows a wavevector resolution of $\Delta k_x=0.03\ \omega_{pe}/c$ and $\Delta k_y=0.25\ \omega_{pe}/c$, respectively.
In order to reduce the level of numerical noise, we use a second order shape function for the macro-particles on the grid. 
The implemented relativistic Boris pusher is relativistically corrected for moderate Lorentz factors ($\gamma \ll 1000$), i.e. for the whole parameter range of interest here.

As a rule the number of macro-particles per cell in PIC code simulations has to be sufficiently large to avoid numerical noise. For the background plasma we choose $N_{bg}=950$ macro-particles per cell and for the electron beam $N_{bm}=50$ macro-particles per cell. The corresponding total numbers of macro-particles used in our simulations are listed in \myreftab{tab:total_number_particles}. In our simulations, the beam-to-background density ratio is hence about $N_{bm}/N_{bg}=1/19$ (or $5.2\%$). This choice results from a compromise between the very dilute beams observed in astrophysical plasma systems as in the solar corona and the computational cost of PIC simulations, which increases considerably the more dilute the beams are. The reason is that denser beams cause a faster relaxation of the distribution function and thus they allow to investigate the instability growth within shorter (computational) time-scales. As a result, typical values higher than $1\%$ of the beam-to-background density ratio are usually used for beam plasma PIC code simulations~\citep[see, e.g.,][]{Thurgood2015,Zhou2020,Reid2018}.
In our case it is specially critical to choose an even higher value because of two other reasons. First, the ECMI develops in cavities, where the (background) density is significantly lower than that in the surroundings. This leads to an even higher beam-to-background density ratio in cavities than elsewhere. This was, perhaps, the reason why most of the previous studies of the ECMI were carried out with beam-to-background density ratios of the order of $5\%$ or even higher~\citep[see, e.g.,][]{Pritchett1984, Lee2011,Zhou2020}.
Second, different from most of previous studies where an extended and homogeneous beam is pushed to repeatedly travel many times throughout the simulation domain because of the periodic boundary conditions, we allow only one pass of the localized beam through the simulation domain along the parallel, inhomogeneous, direction (see details later). 
Therefore, the beam needs to be relaxed to some degree during such one passage, and within the corresponding short time period, for our investigations of plasma wave emission as a consequence of beam plasma instabilities. This in turn requires a high beam-to-background density ratio.

\begin{center}
    \begin{threeparttable}
        \begin{tabular}{c|c|c}
            \hline
            \multirow{2}{*}{Run} &\multicolumn{2}{c}{Number of macro-particles}\\
            \cline{2-3} \\
            & beam electrons & all particles \\
            \hline
            1,2 & \multirow{4}{*}{1,\ 597,\ 398} & 3,\ 987,\ 783,\ 596\\
            3,4 &                                & 1,\ 860,\ 644,\ 780\\
            5,6 &                                & 1,\ 382,\ 435,\ 756\\
            7,8 &                                & 1,\ 342,\ 554,\ 028\\
            \hline
        \end{tabular}
        \caption{Number of macro-particles of beam electrons and of all particles (electrons and protons) used in the simulations.}
        \label{tab:total_number_particles}
    \end{threeparttable} 
\end{center}

\begin{figure}
    \centering
    \includegraphics[width=0.85\textwidth]{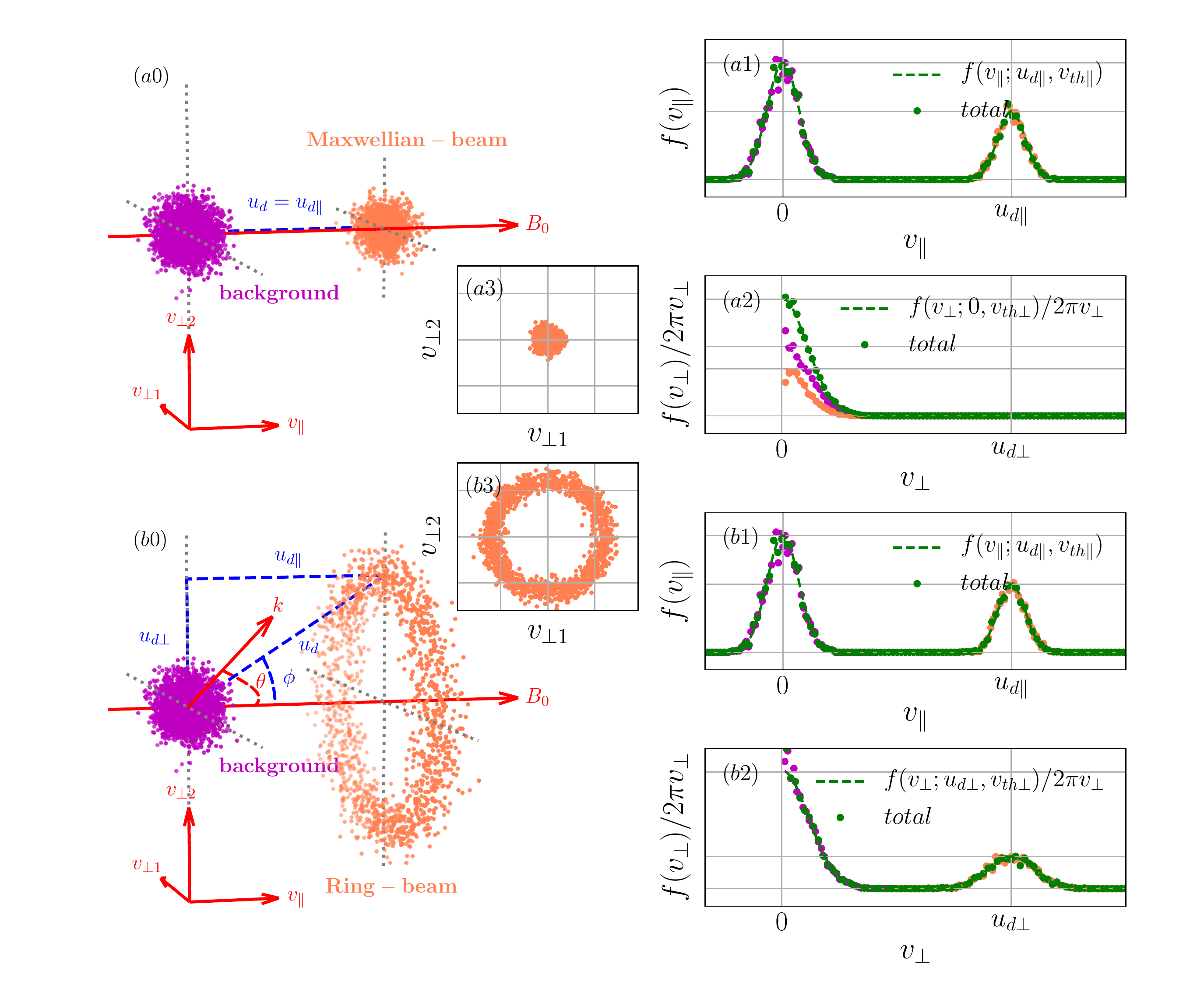}
    \caption[]{Electron distribution of Maxwellian/ring-beam plasma system in the velocity space. (a0, b0) Electron distribution of Maxwellian/ring-beam (coral dots) and background plasma (magenta dots) in the 3D velocity space. Parallel/perpendicular EVDF of Maxwellian-beam (a1, a2) and ring-beam (b1, b2) plasma system. The dashed curves are obtained by the normalized \myrefeq{eq:PDFxvparavperpBackground} and \myrefeq{eq:PDFxvparavperpBeam} in parallel/perpendicular direction respectively. The coloured dots indicate 1D histograms of electrons in the parallel/perpendicular velocity space . (a3, b3) 2D EVDF of Maxwellian/ring-beam plasma system in the $v_{\perp 1}\ vs.\ v_{\perp 2}$ plane respectively.}
    \label{fig:e_vspace_evdf}
\end{figure}

We initialized the beam-plasma by prescribing a particle distribution function $f(\boldsymbol{x},\boldsymbol{v})$ in the phase space $\boldsymbol{x}\times\boldsymbol{v}$. We refer to  a ``homogeneous'' background plasma if $\nabla_{\boldsymbol{x}} f_{bg}=0$ (constant density), and  to an ``inhomogeneous'' background plasma if $\nabla_{\boldsymbol{x}} f_{bg}\neq0$ (inhomogeneous density).
Here the subscript ``bg'' denotes the background plasma and the subscript ``bm'' indicates the beam plasma.
We denote the beam as ``global'' if it is distributed over the whole simulation box~\citep[e.g.,][]{Lee2011}, or ``localized'' if it is limited in space along $x$ or $y$ direction of the simulation box~\citep[e.g., see][]{Sakai2005,Tsiklauri2011b}. 
We always initialize the background plasma in thermal-equilibrium, i.e., by a Maxwellian distribution function in the momentum space. 
As for the beam, we call it Maxwellian beam if its particles obey a Maxwellian distribution function (see~\myreffig{fig:e_vspace_evdf} (a0)), or a ring beam if the particles follow a ring distribution function (see~\myreffig{fig:e_vspace_evdf} (b0)) in the momentum space.
Note that both distribution functions offer a parallel source of free energy, i.e., $v_{\parallel}\cdot\partial f(v_{\parallel})/\partial v_{\parallel}>0$ in $f(v_{\parallel})$ (see \myreffig{fig:e_vspace_evdf}(a1) and \myreffig{fig:e_vspace_evdf}(b1)), but only the ring distribution function can offer a perpendicular source of free energy, i.e., $\partial (f(v_{\perp})/2\pi v_{\perp})/\partial v_{\perp}>0$ in $f(v_{\perp})/2\pi v_{\perp}$ (see \myreffig{fig:e_vspace_evdf}(b2)). The latter is required for the ECME mechanism.

From here on we use the term ``momentum'' for the momentum per unit mass, which is equivalent to the relativistic velocity in its four-vector form, i.e., $p_{\parallel}=v_{\parallel}$, $p_{\perp}=v_{\perp}$. Hence, in the following the velocity components $v_{\parallel}$ and $v_{\perp}$ are in fact mean momenta.
As a result the Lorentz factor is $\gamma=\sqrt{1+(v_{\parallel}^2+v_{\perp}^2)/c^2}$, which corresponds to a particle kinetic energy $E_k=(\gamma-1)m_ec^2$. 

In the phase space $(x,y)\times(v_{\parallel},v_{\perp})$, the particle distribution function of the background plasma is expressed as follows,

\begin{equation}
    f_{bg}\left(x,v_{\parallel},v_{\perp}\right)=n_{bg}(x)\cdot f_{\parallel}\left(v_{\parallel};v_{the}\right)f_{\perp}\left(v_{\perp};v_{the}\right)\label{eq:PDFxvparavperpBackground}
\end{equation}
where
\begin{align}
    n_{bg}(x)&=n_{0,bg}\left[\eta^{-1}+(1-\eta^{-1})\left(\frac{x}{L_0}-1\right)^2\right]\label{eq:PDFxBackground}\\
    f_{\parallel}\left(v_{\parallel};v_{the}\right)&=\frac{1}{\sqrt{2\pi v_{the}^2}}\exp\left(-\frac{v_{\parallel}^2}{2v_{the}^2}\right)\label{eq:PDFvparaBackground}\\
    f_{\perp}\left(v_{\perp};v_{the}\right)&=\frac{1}{v_{the}^2}v_{\perp}\cdot \exp\left(-\frac{v_{\perp}^2}{2v_{the}^2}\right)\label{eq:PDFvperpBackground}
\end{align}
here $n_{0,bg}=N_{bg}(M/\Delta V)=n_0$ is the background number density at $x=0$, $N_{bg}$ is number of macro-particles per cell of the background plasma ($=950$), $M$ is the ratio of physical to numerical particles and $\Delta V$ is the cell volume, $v_{the}$ is the thermal speed of the background electrons.  

\begin{figure}
    \centering
    \includegraphics[width=0.85\textwidth]{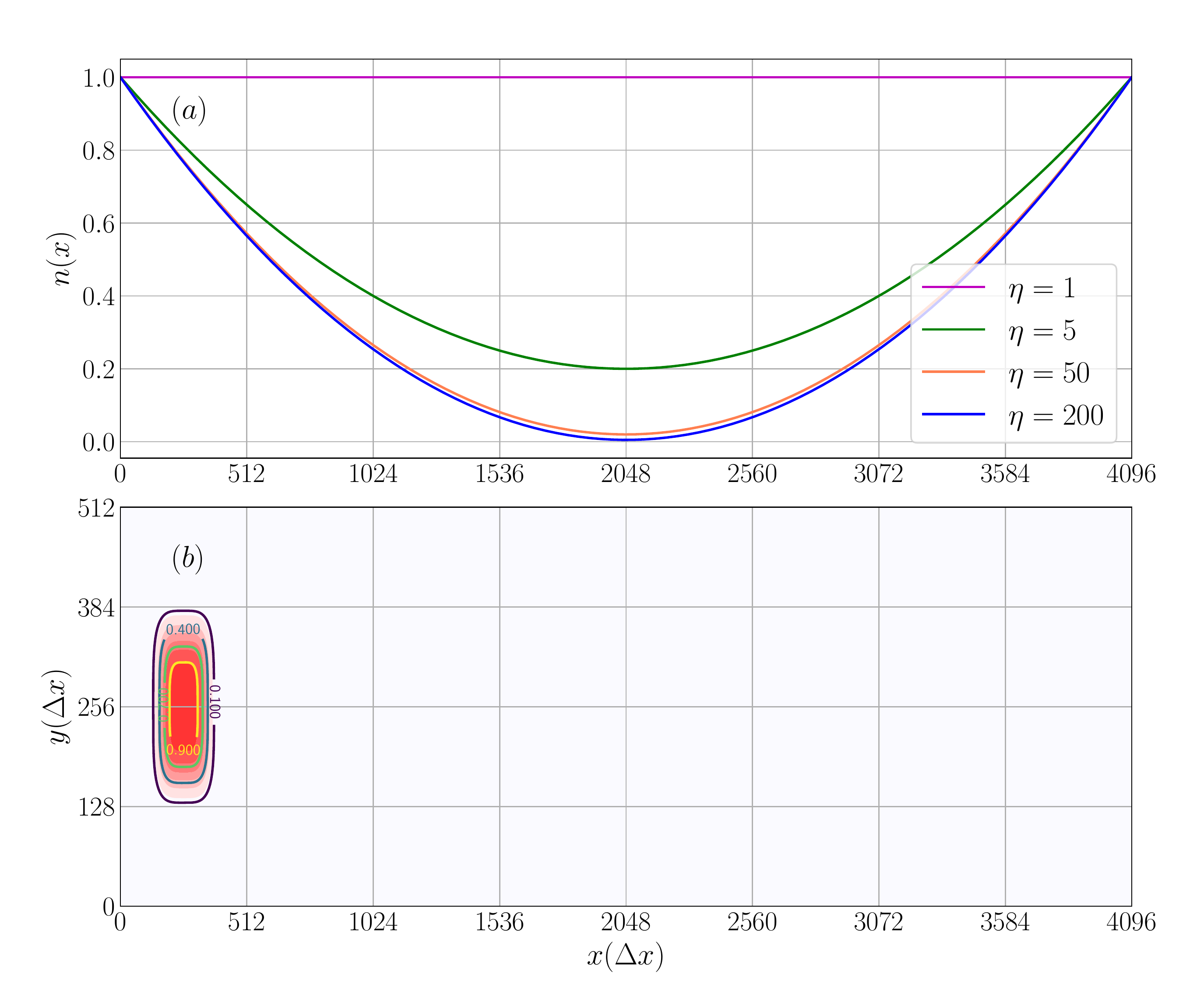}
    \caption[]{(a) Parabolic profile of the normalized background plasma density $n_{bg}(x)/n_{0,bg}$ along the $x-$direction for the homogeneous ($\eta=1$) and  inhomogeneous ($\eta>1$) background plasma. (b) Distribution of the normalized electron density of the localized beam $n_{bm}(x,y)/n_{0,bm}$ in the $x\ vs.\ y$ plane with overlaid contour lines.}
    \label{fig:e_xspace}
\end{figure}

In order to describe a simple density gradient, but at the same time allowing periodic boundary conditions, we use a parabolic density profile (i.e., \myrefeq{eq:PDFxBackground}) for the background plasma. We set $L_0=L_x/2$, thus the parabolic profile is symmetric in the $x$ direction. The density gradient $\eta=n_{max}/n_{min}$, the ratio of maximum to minimum particle number density, describes a density drop from the edge to the centre of the simulation box.
The maximum density is $n_{max}=n_{0}$ at $x=0$ while the minimum density is reached at the centre, i.e., $n_{min}=n_{bg}(x=L_0)=\eta^{-1}n_{0}$.
A density profile with $\eta>1$ represents an inhomogeneous background plasma, while $\eta = 1$ corresponds to the limit case of constant (homogeneous) background density. We used the following values of density gradients in our simulations: $\eta=1,\ 5,\ 50,\ 200$. The corresponding density profiles are shown in ~\myreffig{fig:e_xspace}(a).

We localized the electron beam in the phase space by the following distribution function:
\begin{equation}
    f_{bm}\left(x,y,v_{\parallel},v_{\perp}\right)=n_{bm}(x,y)\cdot f_{\parallel}(v_{\parallel};u_{d\parallel},v_{th\parallel})f_{\perp}(v_{\perp};u_{d\perp},v_{th\perp})\label{eq:PDFxvparavperpBeam}
\end{equation}
where
\begin{align}
    n_{bm}(x,y)& = n_{0,bm}\exp\left[-\left(\frac{x-L_1}{\sigma_1}\right)^n-\left(\frac{y-L_2}{\sigma_2}\right)^n\right]\label{eq:PDFxBeam}\\
    f_{\parallel}(v_{\parallel};u_{d\parallel},v_{th\parallel})&=\frac{1}{\sqrt{2\pi v_{th\parallel}^2}}\exp\left[-\frac{\left(v_{\parallel}-u_{d\parallel}\right)^2}{2v_{th\parallel}^2}\right]\label{eq:PDFvparaBeam}\\
    f_{\perp}(v_{\perp};u_{d\perp},v_{th\perp})&=\frac{1}{C_{\perp}}v_{\perp} \cdot \exp\left[-\frac{\left(v_{\perp}-u_{d\perp}\right)^2}{2v_{th\perp}^2}\right]\label{eq:PDFvperpBeam}
\end{align}
Here the normalization factor $C_{\perp}$ is
\begin{equation}
    C_{\perp}=v_{th\perp}^2 \exp\left(-\frac{u_{d\perp}^2}{2v_{th\perp}^2}\right)+\sqrt{\frac{\pi}{2}}u_{d\perp}v_{th\perp}\left[1+{\rm erf}\left(\frac{u_{d\perp}}{\sqrt{2}v_{th\perp}}\right)\right]\label{eq:normfactorperp}
\end{equation}
where $n_{0,bm}=N_{bm}(M/\Delta V)$ is the 
beam number density at $x=L_1,y=L_2$, $N_{bm}$ is number of macro-particles per cell of the beam ($=50$).
$L_1$ and $L_2$ determine the initial location of the beam, $\sigma_1$ and $\sigma_2$ are the characteristic widths of the localized beam in $x$ and $y$ directions respectively. In our simulations, we set $L_1=150\Delta x,\ L_2=L_y/2=256\Delta x$, $\sigma_1=\sigma_2=100\Delta x$, and $n=4$ (see \myreffig{fig:e_xspace}(b)).  
$v_{th\parallel}$ and $v_{th\perp}$ are the thermal speeds (momenta) of the beam electrons along the parallel and perpendicular directions respectively. In our simulations they are equal, i.e., $v_{th\parallel}=v_{th\perp}=v_{th,bm}$.

Drift velocities of electron beams were deduced from observations of Type III radio bursts as summarized in \citet[][]{Reid2014,Reid2018}.
They were found to be either non-relativistic, $0.2-0.5c$ \citep[][]{Wild1959,Alvarez1973} or mildly relativistic with $>0.6c$ \citep[][]{Poquerusse1994a,Klassen2003}.
For our simulations with either a Maxwellian- or ring- beam, the initial parallel drift speed is $u_{d\parallel}=0.45\ c$.
While the Maxwellian beam has $u_{d\perp}=0$,
the ring-beam simulations have additionally a perpendicular beam drift speed of $u_{d\perp}=0.55\ c$. These drift velocities correspond to a kinetic energy of $E_k=49.36\ keV$ ($\gamma=1.096$) for  the Maxwellian beam and $E_k=115.89\ keV$ ($\gamma=1.23$) for the ring beam.
The thermal speed of the beam electrons is $v_{th,bm}=0.03\ c$, which corresponds to a temperature of $T_{bm}=5.3\times10^{6}\ K$. 

In order to study the influence of the density inhomogeneities of the background plasma on the excitation of plasma instabilities and thus the generation of electromagnetic (radio) waves, we perform eight simulation runs with different parameter configurations.
These parameter sets are summarized in \myreftab{tab:PICSimulationsParameters}.
Note that we mainly vary the background density gradient. 
These parameter sets were chosen to allow both bump-on-tail instability and ECMI, and for the comparative analysis of the resulting waves. 

\begin{center}
    \begin{threeparttable} 
        \begin{tabular}{c|c|c|c|c|c|c|c}
            \hline
            \multirow{2}{*}{Run} &\multicolumn{3}{c|}{Background} &\multicolumn{4}{c}{Beam}\\
            \cline{2-8}
            & \# & $\eta$ &  $v_{the}$ & \# & $u_{d\parallel}$ &  $u_{d\perp}$ & $v_{th,bm}$\\
            \hline
            1 & \multirow{2}{*}{\ \ homogeneous} & \multirow{2}{*}{1}  & \multirow{8}{*}{$0.025c$} & Maxwellian &  \multirow{8}{*}{$0.45c$} & $0$ & \multirow{8}{*}{$0.03c$}\\
            \cline{1-1}
            \cline{5-5}
            \cline{7-7}
            2 &       &   &  & Ring &  & $0.55c$  & \\
            \cline{1-3}
            \cline{5-5}
            \cline{7-7}
            3 & \multirow{6}{*}{inhomogeneous}   & \multirow{2}{*}{5}     &  & Maxwellian & & $0$ &  \\
            \cline{1-1}
            \cline{5-5}
            \cline{7-7}
            4 &  & &   & Ring & & $0.55c$  & \\
            \cline{1-1}
            \cline{3-3}
            \cline{5-5}
            \cline{7-7}
            5 &  & \multirow{2}{*}{50}     &  & Maxwellian &  & $0$ &  \\
            \cline{1-1}
            \cline{5-5}
            \cline{7-7}
            6 &  & &   & Ring & & $0.55c$  & \\
            \cline{1-1}
            \cline{3-3}
            \cline{5-5}
            \cline{7-7}
            7 &  & \multirow{2}{*}{200}    & & Maxwellian &  & $0$ & \\
            \cline{1-1}
            \cline{5-5}
            \cline{7-7}
            8 &  &  &   &Ring &  & $0.55c$ & \\
            \hline
        \end{tabular}
        \caption{Some parameters of our eight simulation runs. Density gradient $\eta=n_{max}/n_{min}$, thermal speed of background electrons $v_{the}$, thermal speed  of beam electrons $v_{th,bm}$, drift speeds $u_{d\parallel}$, $u_{d\perp}$ of electron beam.}
        \label{tab:PICSimulationsParameters}
    \end{threeparttable}
\end{center}

\begin{figure}
    \centering
    \includegraphics[width=0.85\textwidth]{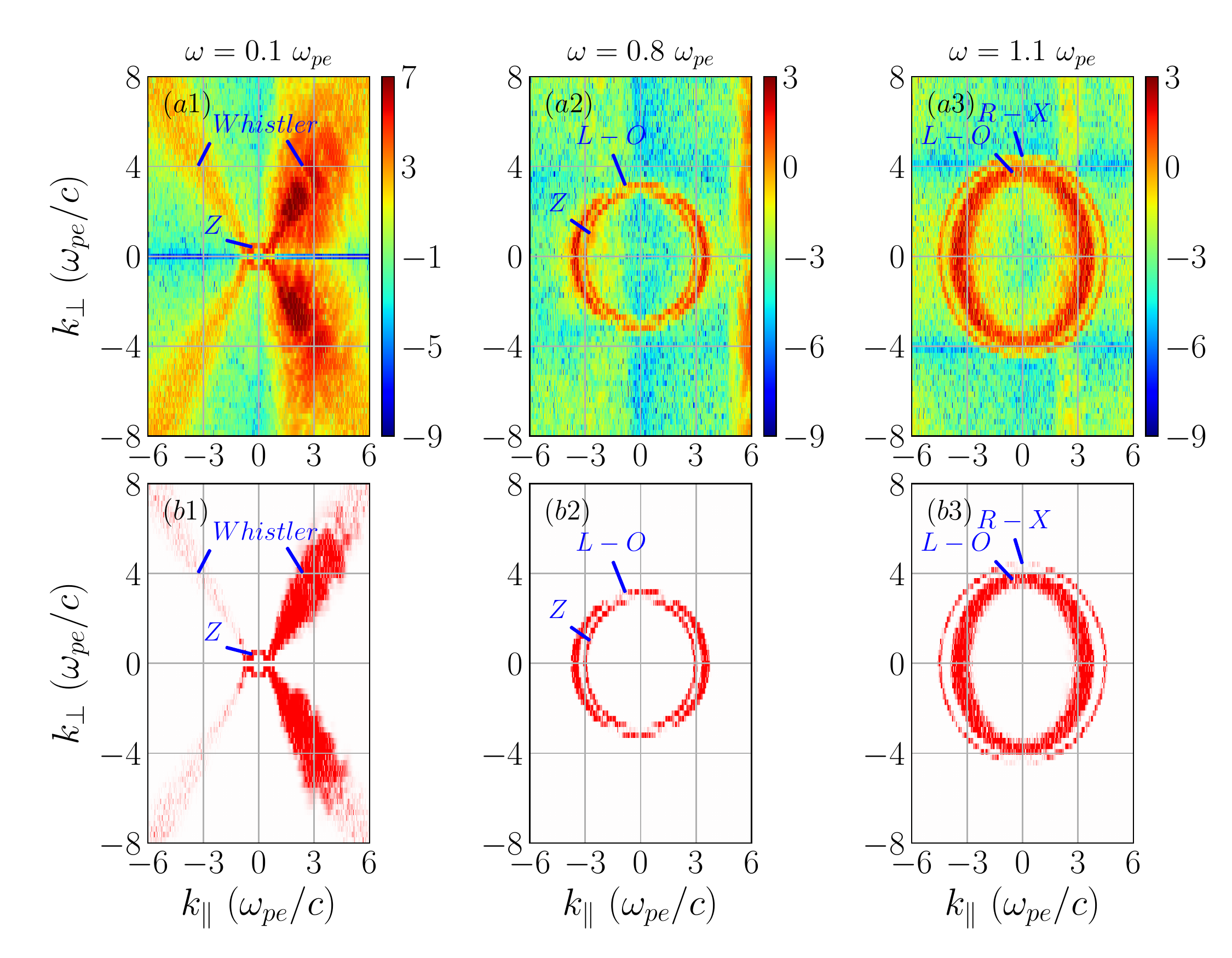}
    \caption[]{(a1-a3) PSD in the $k_{\parallel}\ vs.\ k_{\perp}$ plane for Run1 ($\eta=1$, Maxwellian beam) in the time interval $t=100-164\ \omega_{pe}^{-1}$ at $\omega=0.1\ \omega_{pe},\ 0.8\ \omega_{pe}$ and $1.1\ \omega_{pe}$ respectively. (b1-b3) Extracted PSD of whistler, Z, L-O and R-X modes. The PSDs are evaluated as $\log_{10}\left|E_{y}(k_{\parallel},k_{\perp},\omega)/B_0\right|^2$ and $\log_{10}\left|E_{cy}(k_{\parallel},k_{\perp},\omega)/B_0\right|^2$ separately, with $B_0$ the initial background magnetic field strength.}
    \label{fig:WaveModekw}
\end{figure}

One of our main analysis tools for wave mode diagnostics is the linear wave dispersion theory.
We compare the power spectral density (PSD) derived from electromagnetic fields with the predicted analytical dispersion relation(s). 
The PSD associated to a given wave mode is evaluated by assuming a Gaussian power distribution along its dispersion curve or surface in the Fourier space \citep[e.g.,][]{Wilczek2012a,Comisel2013a}. For example, for the wave mode with dispersion relation $\omega_{c}(k_{\parallel},k_{\perp})$ in the Fourier space $(k_{\parallel},k_{\perp},\omega)$, associated PSD is calculated in the following way:

\begin{align}
    |E_{ci}(k_{\parallel},k_{\perp})|^2&= \int |E_{i}(k_{\parallel},k_{\perp},\omega)|^2\cdot \frac{1}{\sqrt{2\pi\sigma^2}}\exp\left[-\frac{(\omega-\omega_{c}(k_{\parallel},k_{\perp}))^2}{2\sigma^2}\right]d\omega\label{eq:EiWavemodekw}
\end{align}
where $E_i$ ($i=1,2,3$) are the components of the electric field in the Fourier domain. Here $\sigma$ characterizes the frequency broadening of power about the dispersion surface (or curve) of a given wave mode, which is due to thermal effect.
We obtained these dispersion surfaces by solving the linear dispersion relation $\omega_{c}(k_{\parallel},k_{\perp})$ for cold plasmas~\citep[][]{Stix1992}.
In this way one finds the dispersion surfaces (or curves) in the Fourier space $(k_{\parallel},k_{\perp},\omega)$ 
of R-X, L-O, Z and whistler modes.
We found all those predicted wave modes in our simulations:
whistler and Z modes in \myreffig{fig:WaveModekw} (a1); L-O and Z modes in \myreffig{fig:WaveModekw} (a2) and R-X and L-O modes in \myreffig{fig:WaveModekw} (a3). Based on \myrefeq{eq:EiWavemodekw}, PSDs of associated wave modes are extracted from the total PSD separately, e.g., PSDs of whistler and Z modes in \myreffig{fig:WaveModekw} (b1), L-O and Z modes in \myreffig{fig:WaveModekw} (b2) and R-X and L-O modes in \myreffig{fig:WaveModekw} (b3).
The PSDs of Langmuir waves and electron cyclotron waves are extracted in a similar way.

In order to verify that our simulations can correctly reproduce the desired physical instabilities and thus wave emission, we carried out convergence tests by enhancing the grid resolution and the number of macro-particles per cell as discussed in Appendix~\ref{app:CT}. In the following context we discuss our physical results.

\section{Results}\label{sec:results}

\begin{figure}
    \centering
    \includegraphics[width=0.85\textwidth]{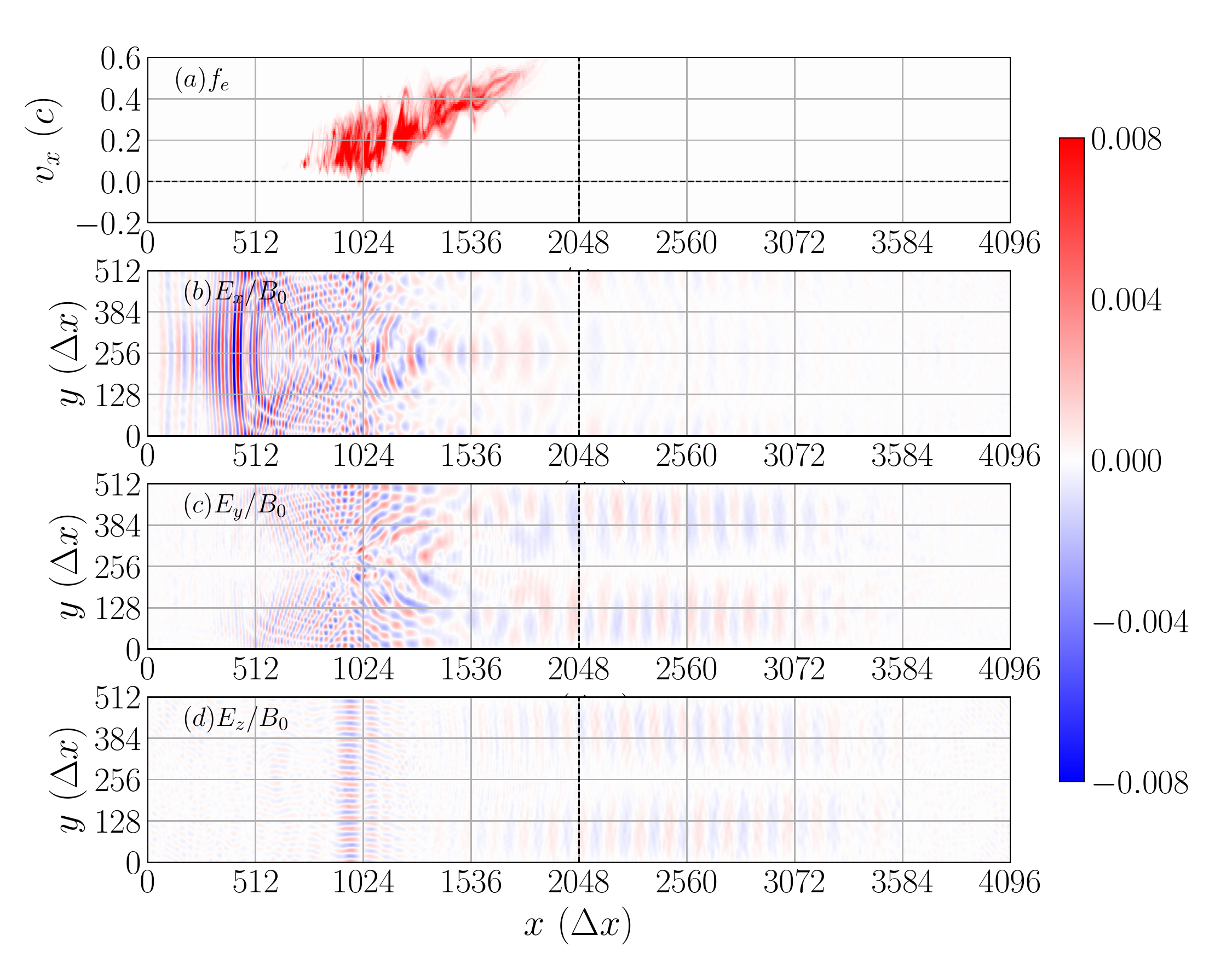}
    \caption[]{(a) Phase-space plot of the electron beam in the $x\ vs.\ v_{x}$ plane. (b-d) Spatial distribution of the electric fields $E_i/B_0\ (i=x,y,z)$ in the $x\ vs.\ y$ plane. All panels show results for Run6 at $t=200\ \omega_{pe}^{-1}$.}
    \label{fig:phase_plot_ne_Exy}
\end{figure}

Our study aims only at investigating the first generation stage of a chain of processes that finally leads to the observed radio emissions.
The processes here analyzed, therefore, take place at short-time scales no more than about
$200\ \omega_{pe}^{-1}$. At this time the beams have not fully relaxed yet but they still emit waves.
\myreffig{fig:phase_plot_ne_Exy} (a) shows that for Run6 the beam is still moving near $x = 1024 \Delta x$ and far from reaching the centre of the simulation box ($x = 2048 \Delta x$) at $t=200\ \omega_{pe}^{-1}$. 
\myreffig{fig:phase_plot_ne_Exy} (b-d) show the spatial structure of three components of electric field $E_i\ (i=x,y,z)$ at $t=200\ \omega_{pe}^{-1}$.
From them we find that waves, propagating ahead the beam, already crossed the centre of the simulation box, but they are not reflected back, yet.
Other simulations exhibit similar phenomena. This allows to conclude that we can practically neglect wave reflections and mode conversion effects.
Those processes might take place after the beam moves beyond the centre and into the right half of the simulation domain, in which region the density increases since we for technical reasons set up a parabolic density profile according to~\myrefeq{eq:PDFxBackground}.

According to the discussion before, it is expected to observe enhanced Langmuir waves  in all the simulations with Maxwellian beam and ring beam, because their distribution functions offer a parallel source of free energy  (positive parallel gradients in their EVDFs). This can cause beam plasma (or bump-on-tail) instabilities (see \myreffig{fig:e_vspace_evdf}(a1, b1)).
For electron cyclotron waves, the situation is different. Even though the Maxwellian beam is unable to offer a source of free energy (positive gradient in the perpendicular EVDF, see \myreffig{fig:e_vspace_evdf}(a2)) that causes the electron cyclotron maser (ECM) instabilities, electron cyclotron waves are actually observed due to electron cyclotron resonances (ECRs). On the other hand, for the ring beam, besides electron cyclotron waves caused by ECRs, electron cyclotron waves produced by ECM instabilities are observed because a perpendicular source of free energy for the ECME is available (see \myreffig{fig:e_vspace_evdf}(b2)).

In the following we first discuss the energy transfer from the kinetic energy of the electron beam into other forms of energy of both plasma and electromagnetic fields (and thus of radio waves). Then we present our results regarding fundamental and harmonics of Langmuir waves and electron cyclotron waves. And finally we show a Doppler frequency shift phenomena happening in the electron cyclotron resonance region.

\subsection{Conversion of kinetic energy}\label{sec:conversion_energy}

\begin{figure}
    \centering
    \includegraphics[width=0.85\textwidth]{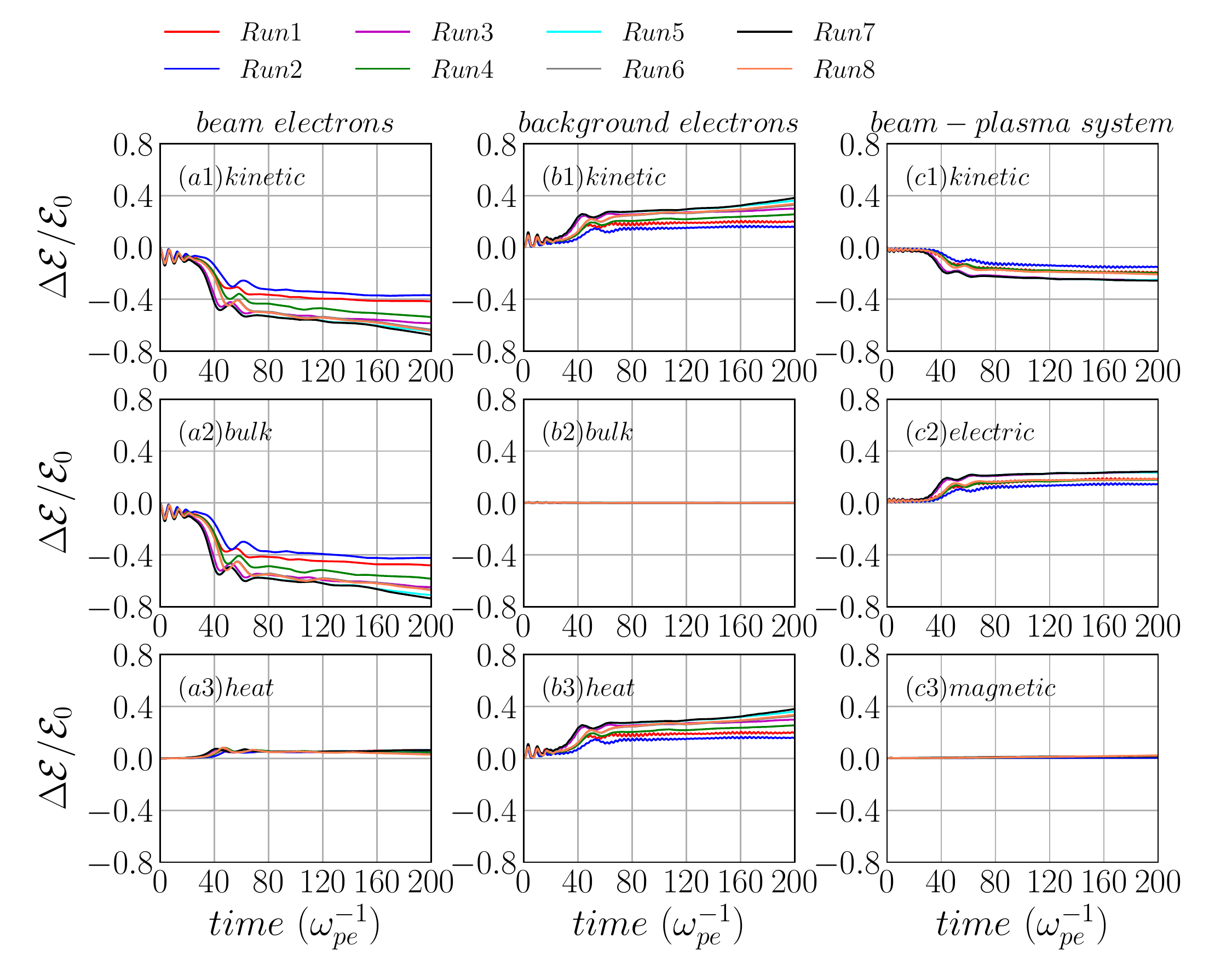}
    \caption[]{Temporal energy variations for all simulations. (a1-a3): Variation of total kinetic energy, bulk flow energy and thermal energy of beam electrons respectively, (b1-b3): same for background electrons, (c1-c3): variations of the total kinetic energy, electric energy and magnetic energy of the beam-plasma system. All the energy variations are normalized by the initial bulk flow kinetic energy of the electron beam $\mathcal{E}_0$ of Run1.}
    \label{fig:Energy_Kinetic_E}
\end{figure}

Before discussing the generation of Langmuir and electron cyclotron waves, let us consider the temporal variation of the kinetic energy of the beam electrons (see \myreffig{fig:Energy_Kinetic_E} (a1-a3)) and of the background electrons (see \myreffig{fig:Energy_Kinetic_E} (b1-b3)). Here the total kinetic energy is calculated by $\mathcal{E}_t=\frac{1}{2}m_e\overline{v^2}\cdot N_e\cdot M$, the bulk flow kinetic energy as $\mathcal{E}_b=\frac{1}{2}m_e\overline{v}^2\cdot N_e\cdot M$, and the electron thermal energy (heat) by $\mathcal{E}_{th}=\mathcal{E}_t-\mathcal{E}_b$. The averages are over all the particles, e.g., $\overline{v}=\left(\displaystyle\sum_i v_i\right)/N_e$, where $v_i=\sqrt{v_{ix}^2 + v_{iy}^2 + v_{iz}^2}$ is the speed of the $i$-th particle. $N_e$ is the number of macro-electrons of the beam or background plasma and $M$ is the ratio of physical to numerical particles as defined before. 
The total electric field energy is calculated as 
$\mathcal{E}_{E}=1/(8\pi)\cdot\displaystyle\sum_{i=1}^{N_x} \sum_{j=1}^{N_y} E_{ij}^2\Delta V$, where $\Delta V=\Delta x\cdot \Delta x$ is the cell area (or ``volume'') for a 2D simulation. The total magnetic energy is calculated in a similar fashion by replacing electric field $E$ with magnetic field $B$.

In order to compare the electron energy gain or loss, we evaluate the normalized variation of each kind of energy as $\Delta \mathcal{E}(t)/\mathcal{E}_0=\left(\mathcal{E}(t)-\mathcal{E}(t=0)\right)/\mathcal{E}_0$. All energy variations are normalized by the corresponding initial bulk flow kinetic energy of the electron beam $\mathcal{E}_0=\frac{1}{2}m_eu_{d\parallel}^2\cdot N_{e,bm}\cdot M$. Since the number of macro-electrons of the beam in all PIC-code simulations is unchanged, and the electron beams are initialized with the same parallel drift speed $u_{d\parallel}$, as a result, the initial energy $\mathcal{E}_0$ is the same for all simulations.
\myreffig{fig:Energy_Kinetic_E} (c1-c3) also shows evolutions of the total kinetic energy, the electric and magnetic energy fluctuations of beam-plasma systems.
The results displayed in \myreffig{fig:Energy_Kinetic_E} show:

(1) most of the bulk flow kinetic energy of the beam electrons is transferred into thermal energy of the background electrons while only a small part of it is converted into electric field energy;

(2) the background electrons are heated while their bulk kinetic energy remains nearly zero;

(3) the electric energy increases at the expense of the bulk flow kinetic energy of the beam electrons.

Comparing the energy variations obtained for different parameter regimes displayed in \myreffig{fig:Energy_Kinetic_E}, we found that the density gradient of the background plasma significantly influences the energy conversion process for both Maxwellian and ring beam: the larger the density gradient $\eta$ is, the more efficient the kinetic energy is to convert into other form of energies. Comparing energy variations for different cases with same density gradient $\eta$, e.g. Run1 and Run2, or Run3 and Run4, or Run5 and Run6, we found that when $\eta\le50$ the energy conversion of the Maxwellian beam case is more efficient than that of the ring beam case. When  $\eta> 50$, however, the energy conversion in ring-beam cases is more efficient than that in Maxwellian-beam cases, e.g., Run7 and Run8.

\subsection{Langmuir waves}

Our simulations revealed the generation of electrostatic Langmuir waves as well as of electromagnetic electron cyclotron waves. Let us first discuss the formation of Langmuir waves and their consequences for possible radio wave emissions. 

Langmuir waves are generated by the bump-on-tail instabilities for both Maxwellian-beam (Runs 1, 3, 5, 7) and ring-beam (Runs 2, 4, 6, 8) cases. The results of our simulations show that the bump-on-tail EVDF of Maxwellian- and ring-beam plasma systems with same density gradient have nearly the same influence on the generation of Langmuir waves. In this part, only results associated to Maxwellian beams are discussed in detail, because the same results also hold for the ring beam cases.

\begin{figure}
    \centering
    \includegraphics[width=0.85\textwidth]{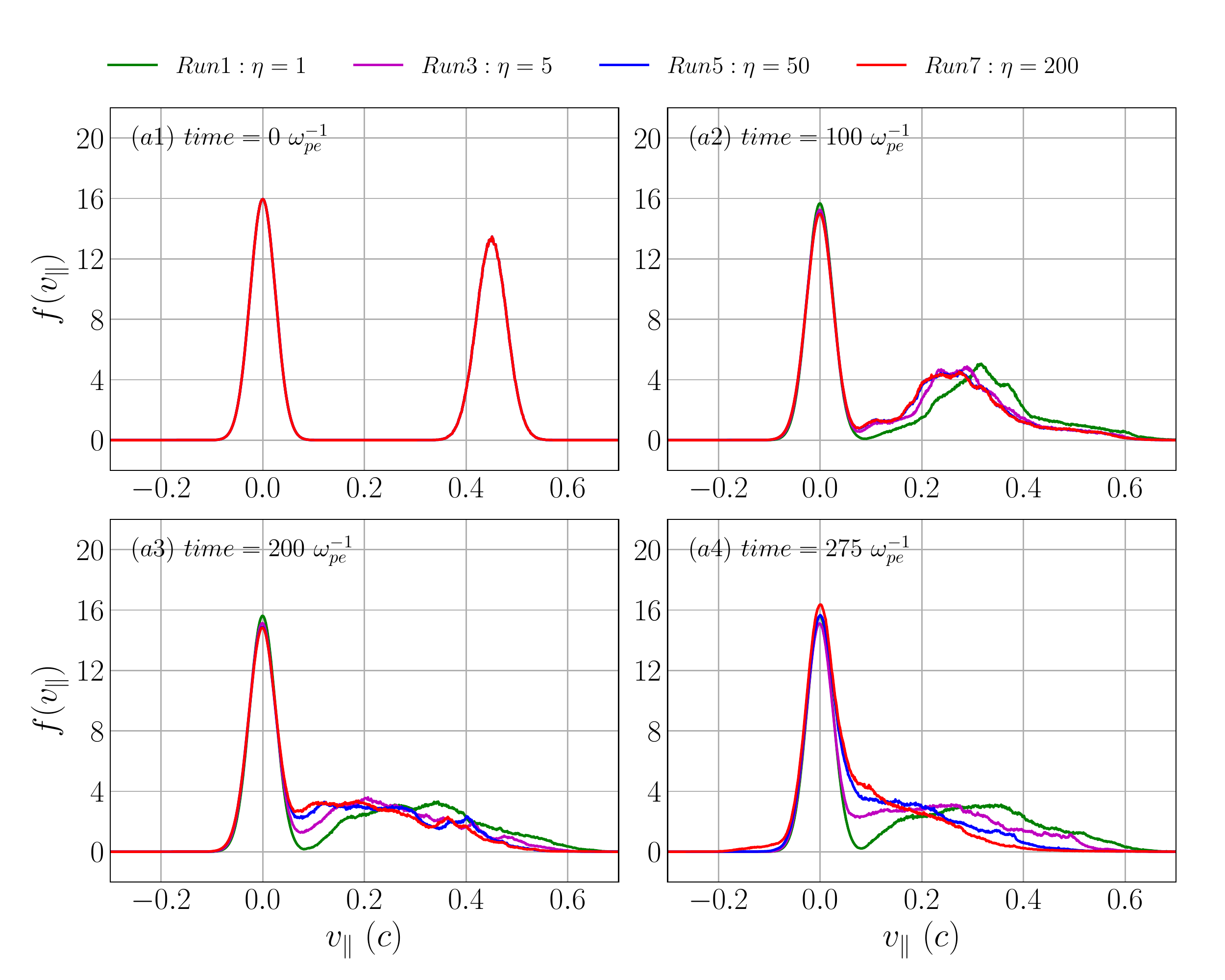}
    \caption[]{Temporal evolution of the parallel EVDF for Runs 1, 3, 5, 7 at time $t=0,\ 100,\ 200,\ 275\ \omega_{pe}^{-1}$, respectively. 
	The background (centred at $v_{\parallel}=0$) and beam  (centred at $u_{d\parallel}$) EVDFs are probability density functions but independently normalized for visualization purposes, so that their respective integrations in the whole $v_{\parallel}$ space yield 1.}
    \label{fig:paraprofile}
\end{figure}

How do the electron populations relax in the parallel velocity space?  \myreffig{fig:paraprofile} depicts the temporal evolution of the bump-on-tail EVDF $f(v_{\parallel})$ of Maxwellian beam-plasma cases (i.e., Runs 1, 3, 5, 7) with increasing density gradients (i.e., $\eta=1,\ 5,\ 50,\ 200$) at $t=0, 100, 200$ and $275 \ \omega_{pe}^{-1}$, respectively. Since the beam density is much smaller than the background density, we normalize the 1D EVDFs of background and beam independently in order to visualize them in the same plot. 
Note that $t=275\ \omega_{pe}^{-1}$ is beyond the maximum time $t=200\ \omega_{pe}^{-1}$ of the duration in which we analyzed plasma waves. 

\begin{figure}
    \centering
    \includegraphics[width=0.85\textwidth]{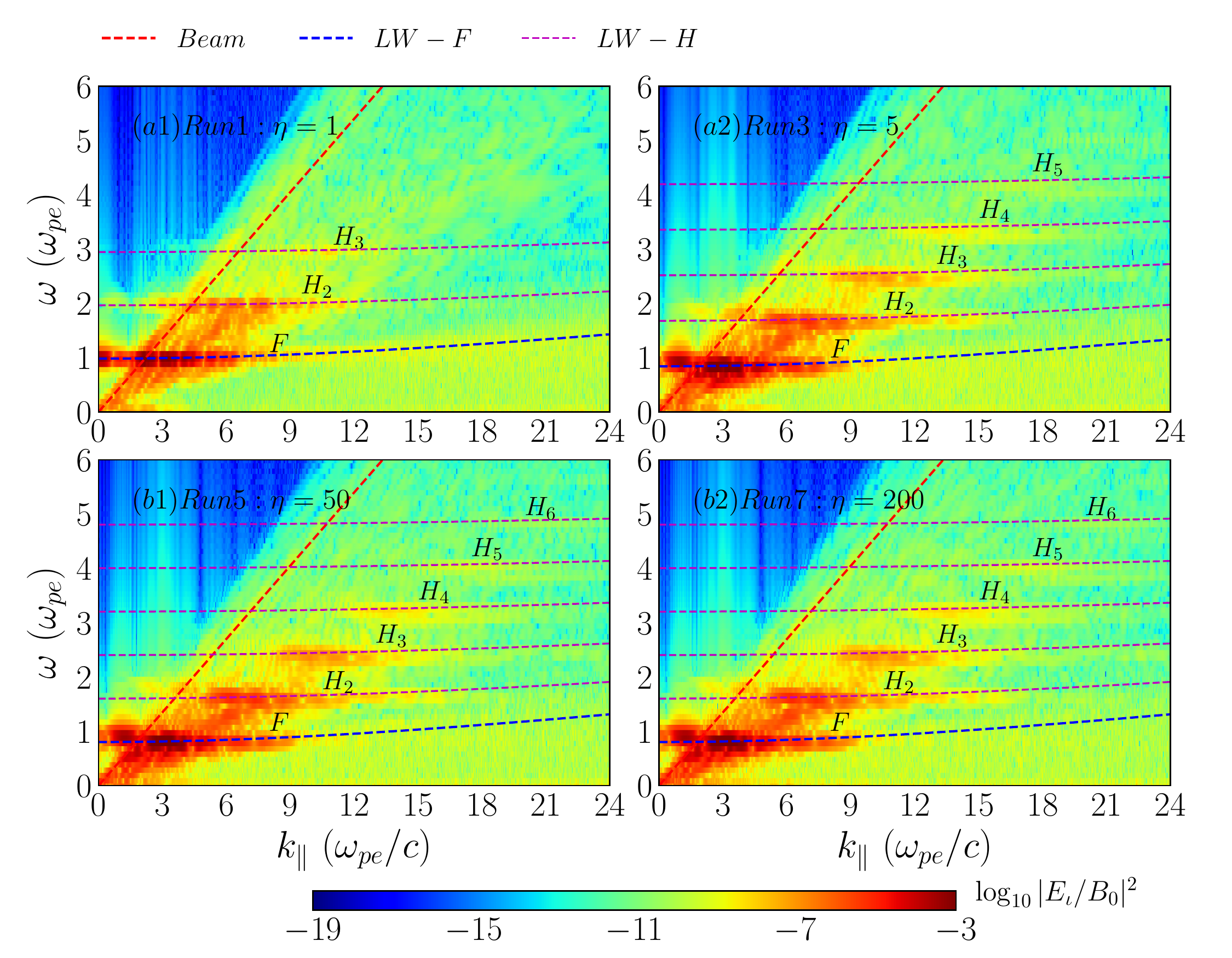}
    \caption[]{Illustration of fundamental (F) and harmonics (H) of Langmuir wave (LW) modes. PSD in the $\ k_{\parallel}\ vs.\ \omega$ plane for Runs 1, 3, 5, 7 (Maxwellian beam) during $t=100-164\ \omega_{pe}^{-1}$ are displayed. The PSD is evaluated as $\log_{10}|E_{\iota}/B_0|^2$. Dispersion relation curves of beam mode (red dashed line), fundamental (blue dashed curve) and harmonics (magenta dashed curves) of Langmuir waves are overlaid.}
    \label{fig:LangmuirwavesHF}
\end{figure}

The beam of Run7 (red curves in~\myreffig{fig:paraprofile}), for example, is relaxing to form a characteristic plateau between $t=100-200\ \omega_{pe}^{-1}$, while the plateau is already formed around $t=200-275\ \omega_{pe}^{-1}$.
As the density gradient of background plasma increases (from Run1 to Run 7), this beam relaxation process occurs faster. This behaviour is correlated to the more efficient transfer of the bulk flow kinetic energy of the electron beam into energy of plasma waves (see \myreffig{fig:Energy_Kinetic_E}).
Note that the beam EVDF has not yet fully relaxed in each simulation while harmonics of Langmuir waves are generated even at earlier stages of the beam relaxation  (see \myreffig{fig:LangmuirwavesHF}). This means that the free energy released from electron beam causes beam plasma instabilities and thus generate Langmuir waves and their harmonics from the very beginning of the beam relaxation. For example, \myreffig{fig:LangmuirwavesHF} shows that harmonics of Langmuir waves are generated between $t=100-164\ \omega_{pe}^{-1}$. 

Note that a full beam relaxation, in particular for a small relative density gradient, would require a much larger simulation box. This would be computationally very demanding.
In more precise terms, let us consider the relative gradient $G=(n_{max}-n_{min})/(L_x/2)=n_{max}(1-\eta^{-1})/(L_x/2)$, where $L_x$ the simulation box size along the $x$ direction, so that the denominator is the distance between the location of the maximum density at the left boundary to the minimum density at the center of the simulation box. Note that the quantity $\eta=n_{max}/n_{min}$ in the numerator is the density gradient defined before.
In general, in order to keep $G$ constant, a larger simulation domain $L_x$ would require a larger $\eta$. However, in order to avoid numerical artifacts, like numerical heating and strong electric field fluctuations due to charge separation, the minimum number of particles per cell $N_{bg}\cdot\eta^{-1}$ should not be smaller than approximately 5-10 particles per cell, assuming a constant macro-factor or weight (ratio of physical to numerical particles) for each macro-particle. This situation could be partially avoided by using a variable macro-factor so that regions with low physical density can be represented by more numerical particles, but we do not utilize such a feature in the present simulations.  On the other hand, in our simulations the maximum number of particles per cell is constrained to be smaller than 1000 due to computational reasons. Thus the upper bound of our density gradient is $\eta\le N_{bg}/5\approx 200$.
Therefore, if we want to have the same gradient $G$ but we are constrained by an upper bound on $\eta$, we can only choose a maximum simulation domain size $L_x$ for the given $\eta$. A larger simulation domain size $L_x$ for a given $\eta$ would imply a smaller gradient $G$.

In the following we will only concentrate on the electric field $\boldsymbol{E}$.  We refer to $E_{\iota},\ E_{\tau},\ E$ as the longitudinal, transverse and total electric field in Fourier space, i.e.,

\begin{equation}
    \left\{
    \begin{aligned}
        E_{\iota}&=E_x\\
        E_{\tau}&=\sqrt{E_y^2+E_z^2}\\
        E&=\sqrt{E_x^2+E_y^2+E_z^2}
    \end{aligned}
    \right.
\end{equation}
The power spectral density (PSD) of each electric field component is simply evaluated by squaring each of those components and normalizing properly.

For a time window spanning of $64\ \omega_{pe}^{-1}$, the variation of the local plasma frequency along the beam trajectory can be neglected.
This implies that the the effective plasma frequency can be treated as a constant when performing the Fast Fourier transform (FFT) on the electromagnetic fields. 
\myreffig{fig:LangmuirwavesHF} shows the resulting PSD derived from longitudinal electric field $E_{\iota}$ for Runs 1, 3, 5 and 7 (Maxwellian beams) in the $(k_{\parallel},\omega)$ domain during $t=100-164\ \ \omega_{pe}^{-1}$ as well as the corresponding analytical dispersion curves of fundamental Langmuir waves and up to their sixth harmonic. A PSD enhancement along these dispersion relation curves indicates that those harmonics of Langmuir waves are indeed generated. 

The dispersion relation of the fundamental (F) Langmuir waves $L$ can be expressed by the following expression: 

\begin{align}
    \omega_{L}=\sqrt{\omega_{loc}^2+3v_{the}^2k^2}\label{eq:LangmuirF}
\end{align}
here $\omega_{loc}$ is the effective local plasma frequency, which is dependent on the location of electron beam as it propagates through the background plasma. $v_{the}$ is the electron thermal speed of background plasma. Note that in a homogeneous background plasma the local plasma frequency remains constant, i.e., $\omega_{loc}=\omega_{pe}$, then \myrefeq{eq:LangmuirF} becomes the standard Bohm-Gross dispersion relation of fundamental Langmuir waves~\citep[][]{Yoon2003a,Melrose2017a}.

For the higher harmonic of Langmuir waves,
we generalized \myrefeq{eq:LangmuirF} to the dispersion relation in the following form, says
\begin{align}
    \omega_{L_n}(k)=\sqrt{n^2\omega_{loc}^2+3v_{the}^2k^2}\label{eq:LangmuirFH}
\end{align}
here $L_n\ (n\ge 2)$ denotes the $n$-$th$ harmonic of Langmuir waves. This is just an empirical relation that fits well with harmonics of the Langmuir waves in our simulations.

In order to compare our formula \myrefeq{eq:LangmuirFH} with previous works,
we can take its long-wavelength limit, i.e., $\omega_{loc}^2\gg 3v_{the}^2k^2$, so that:
\begin{align}
    \omega_{L_n}(k)\approx n\omega_{loc}\left(1+\frac{3}{2n^2}\lambda_{D}^2k^2\right)\label{eq:LangmuirFH_approx}
\end{align}
this is similar to but not the same as previously reported results~\citep[e.g.,][]{Yoon2000b,Gaelzer2002,Yoon2003a}. For the second harmonic ($n=2$), \myrefeq{eq:LangmuirFH} as well as \myrefeq{eq:LangmuirFH_approx} are equivalent to Eq.(34) of~\citet[][]{Yoon2000b}, while for higher harmonics of Langmuir waves, i.e., $n>2$, \myrefeq{eq:LangmuirFH} and \myrefeq{eq:LangmuirFH_approx} are more accurate due to a factor $1/n$ in their second term within the parenthesis on the right-hand-side of \myrefeq{eq:LangmuirFH_approx} (comparing also to Eq.(18) in~\citet[][]{Yoon2003a}).
This discrepancy with previous theoretical works is an indication of a different physical process causing those harmonic waves in our simulations.
A theoretical investigation of the dispersion relation of harmonics of Langmuir waves based on kinetic theory is needed but beyond the goal of this study. 
Note that \myrefeq{eq:LangmuirFH_approx} is still valid for $\omega_{loc}^2\approx 3v_{the}^2k^2$, as confirmed by our simulations.

As the electron beam propagates through the background plasma, the local plasma frequency changes due to the density gradient. \myreftab{tab:effective_wloc} shows an algorithm to estimate the effective local plasma frequency $\omega_{loc}$ based on the power spectrum derived from the longitudinal electric field $E_{\iota}$. 
Once the effective local plasma frequency is determined, the dispersion relation can be solved for the harmonics of Langmuir waves by using \myrefeq{eq:LangmuirFH}.

\begin{table}
    \begin{center}
        \begin{tabular}{p{1.035cm}p{10cm}}
            Step 1. & Integrate PSD of the electrostatic field $E_{\iota}\left(\omega,k_{\parallel},k_{\perp}=0\right)$ over $k_{\parallel}$ to obtain its power spectrum $\mathcal{P}(\omega)$ according to \myrefeq{eq:EnergyLongitudinalEl};\\
            Step 2. & find the characteristic frequencies by $k_{\parallel}$-locations of the local maximum values of $\mathcal{P}(\omega)$;\\
            Step 3. & divide any characteristic frequency by its harmonic number $n$ to get the effective local plasma frequency $\omega_{loc}$. 
        \end{tabular}
        \caption{Algorithm to determine the effective local plasma frequency $\omega_{loc}$}
        \label{tab:effective_wloc}
    \end{center}
\end{table}

\myreffig{fig:LangmuirwavesHF} shows that the steeper the background density gradient (larger $\eta$) is, the more Langmuir harmonics appear.
For Run1 ($\eta=1$, see \myreffig{fig:LangmuirwavesHF} (a1)) only three harmonics are generated, while for Run5 ($\eta=50$, see \myreffig{fig:LangmuirwavesHF} (b1)) at least six harmonics are  generated. As~\myreffig{fig:LangmuirwavesHF} shows, the wavenumber $k_{\parallel}$ of the power spectrum peak of each harmonic mode also increases with the harmonic order $n$.

In order to quantitatively analyze the frequencies of harmonic Langmuir waves, we calculate the power spectrum in the frequency domain by integrating the PSD of the electrostatic (longitudinal) electric field over $k_{\parallel}$, i.e.,

\begin{equation}
    \mathcal{P}(\omega)=\int |E_{\iota}\left(\omega,k_{\parallel},k_{\perp}=0\right)|^2 dk_{\parallel}
    \label{eq:EnergyLongitudinalEl}
\end{equation}

\begin{figure}
    \centering
    \includegraphics[width=0.85\textwidth]{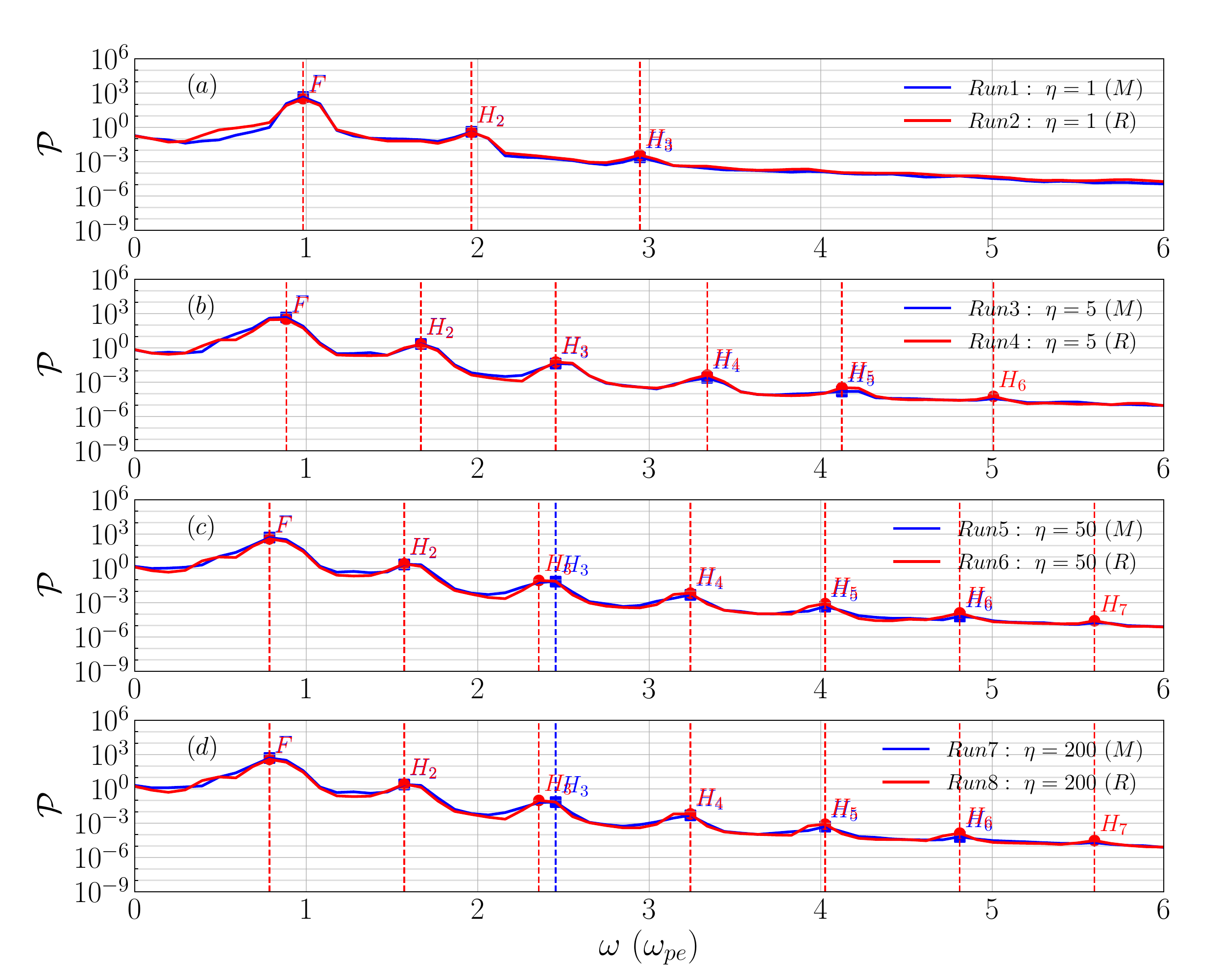}
    \caption[]{Power spectrum $\mathcal{P}(\omega)$ derived from longitudinal electric field $E_{\iota}$ for all simulations in $t=100-164\ \omega_{pe}^{-1}$. The vertical dashed lines denote characteristic frequencies of the fundamental and harmonics of Langmuir waves. The unit of the spectrum is $statV^2\cdot cm^{-2}\cdot s$. Here ``M'' indicates a Maxwellian beam, while ``R'' denotes a ring beam.}
    \label{fig:Energy_LangmuirwavesHF}
\end{figure}

\myreffig{fig:Energy_LangmuirwavesHF} shows the power spectra  $\mathcal{P}(\omega)$ of all simulations 
during the time-period $t=100-164\ \omega_{pe}^{-1}$. We conclude that:

(1) for beam-plasma systems with the same background density gradient $\eta$, we observe practically the same power spectra, in particular at the characteristic frequencies of Langmuir waves and their harmonics. It is because we initialized both Maxwellian- and ring- beams with the same parallel drift speed $u_{d\parallel}$ and thus the same bump-on-tail EVDF;

(2) for beam-plasma systems with different background density gradients, as density gradients $\eta$ increase, the characteristic frequencies of fundamental and harmonics of Langmuir waves shift
to lower frequencies. This is because for larger density gradients $\eta$, the effective local plasma frequency at a beam location gets smaller.

(3) as $\eta$ increases, more harmonics of Langmuir waves are generated. For example, for Maxwellian-beam plasma system Run1 ($\eta=1:F,\ H_2,\ H_3$, see \myreffig{fig:Energy_LangmuirwavesHF}(a)), Run3 ($\eta=5:F,\ H_2,\ \dots,\ H_6$, see \myreffig{fig:Energy_LangmuirwavesHF}(b)), Run5 and Run7 ($\eta=50$ and $200: F,\ H_2,\ \dots,\ H_7$, see \myreffig{fig:Energy_LangmuirwavesHF}(c,d)).

(4) The local maximum power of each harmonic of Langmuir waves significantly decreases as the harmonic number $n$ increases in each beam-plasma system.

(5) Nearly no difference is found between the power spectra of the beam-plasma systems with $\eta=50$ and $\eta=200$ (i.e., Runs 5,\ 6,\ 7,\ 8, see \myreffig{fig:Energy_LangmuirwavesHF}(c,d)). This means that when $\eta\ge 50$, background density gradients affect the Langmuir waves to nearly the same extent.

\myreffig{fig:Energy_LangmuirwavesHF} shows that characteristic frequencies of the third harmonic of Run5 and Run6 (see \myreffig{fig:Energy_LangmuirwavesHF}(c)), and of Run7 and Run8 (see \myreffig{fig:Energy_LangmuirwavesHF}(d)) slightly deviate from each other. This might be due to the frequency resolution $\Delta \omega$ of the FFT. Such deviation is expected to disappear when performing FFT on a longer time series  allowing a higher frequency resolution.

The characteristic frequencies and wavenumbers of fundamental and harmonics of Langmuir waves in our simulations satisfy the wave-wave conditions, also known as beat or Manley-Rowe conditions~\citep[e.g., see Eq.~(1) in][]{Melrose2017a}.
Indeed, by virtue of the dispersion relation~\myrefeq{eq:LangmuirFH_approx}, one can prove the beat condition of the frequencies $\omega_{L_{n-1}}+\omega_{L} \approx \omega_{L_n}$, where $n$ is the harmonic number. 
The wavenumber of each harmonic of Langmuir waves also satisfy their corresponding beat condition $k_{n}=nk_{1}$. $k_n$ corresponds to the $k_{\parallel}$-location of the PSD peak along the dispersion relation curve of the $n$-th harmonic of Langmuir waves. The method to determine such $k_n$ will be explained below.
This provides evidence that the most likely generation mechanism of harmonics of Langmuir waves in our simulations is the coalescence of beam-generated (fundamental) Langmuir waves $L$ and adjacent Langmuir harmonic $L_{n-1}$, which leads to the $n$-th harmonic of Langmuir waves, i.e.,
\begin{equation}
    L_{n-1}+L\to L_n\label{eq:new_mechanism_LHs}
\end{equation}
here $n\ge 2$. In particular, if $n=2$, ~\myrefeq{eq:new_mechanism_LHs} indicates that the second harmonic of Langmuir waves is caused by a coalescence of beam-generated Langmuir waves $L$.
This mechanism was already predicted by previous theoretical studies~\citep[e.g., see][]{Yoon2003a,Yi2007,Rhee2009}. It is important to emphasize that the specific harmonic wave dispersion relation of those previous works is different from ours (see \myrefeq{eq:LangmuirFH}), probably because the underlying physical processes are not exactly the same.
Nevertheless, the harmonics of Langmuir waves observed in our simulations also satisfy the coalescence process \myrefeq{eq:new_mechanism_LHs}.

To analyze the characteristic wavenumbers $k_n\ (n=1,2,3\dots)$ of harmonics of Langmuir waves, we calculate the power distribution $\mathcal{P}_i(k_{\parallel})$ of each harmonic based on \myrefeq{eq:EiWavemodekw} after appropriately ruling out the influence of the beam mode:

\begin{align}
    \mathcal{P}_i(k_{\parallel})=\int{\frac{1}{\sqrt{2\pi}\sigma}\exp\left[-\frac{(\omega-\omega_{L_i})^2}{2\sigma^2}\right]\cdot |E_l(k_{\parallel},k_{\perp}=0,\omega)|^2d\omega}\label{eq:power_Langmuir}
\end{align}
here dispersion relation $\omega_{L_i}$ is determined by \myrefeq{eq:LangmuirFH}. Then $k_n$ of each Langmuir wave harmonic
is determined by the wavenumber $k_{\parallel}$ of the local maximum of $\mathcal{P}(k_{\parallel})$.

\begin{figure}
    \centering
    \includegraphics[width=0.85\textwidth]{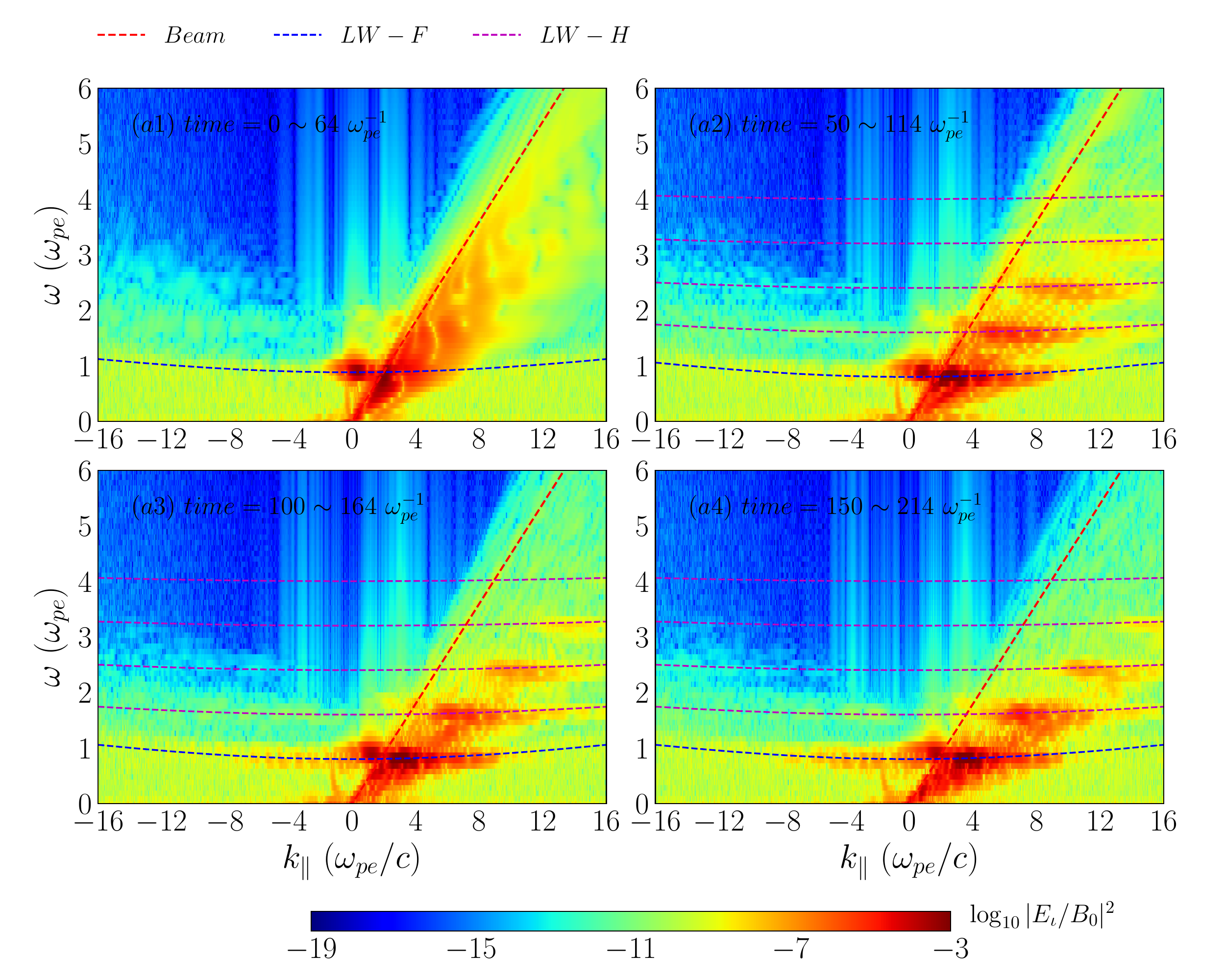}
    \caption[]{PSD in the $k_{\parallel}\ vs.\ \omega$ plane for Run5 ($\eta=50$, Maxwellian beam) in four consecutive time periods $t=0-64,\ 50-114,\ 100-164,\ 150-214\ \omega_{pe}^{-1}$. The PSD is evaluated by $\log_{10}|E_{\iota}/B_0|^2$. Dispersion relation curves of beam mode (red dashed lines), fundamental (blue dashed curve) and harmonics (magenta dashed curves) of Langmuir waves are overlaid.}
    \label{fig:PSD_El_Temporal}
\end{figure}

Note that this characteristic wavenumber $k_n$ is time-dependent. \myreffig{fig:PSD_El_Temporal} shows the temporal evolution of the PSD of longitudinal electric field $E_{\iota}$ of Run5 ($\eta=50$) at time periods $0-64,\ 50-114,\ 100-164,\ 150-214\ \omega_{pe}^{-1}$, respectively. 
As time evolves, the $k_{\parallel}$-location of the PSD peak along its dispersion relation curve of each Langmuir harmonic moves to higher wavenumber region.

\begin{figure}
    \centering
    \includegraphics[width=0.85\textwidth]{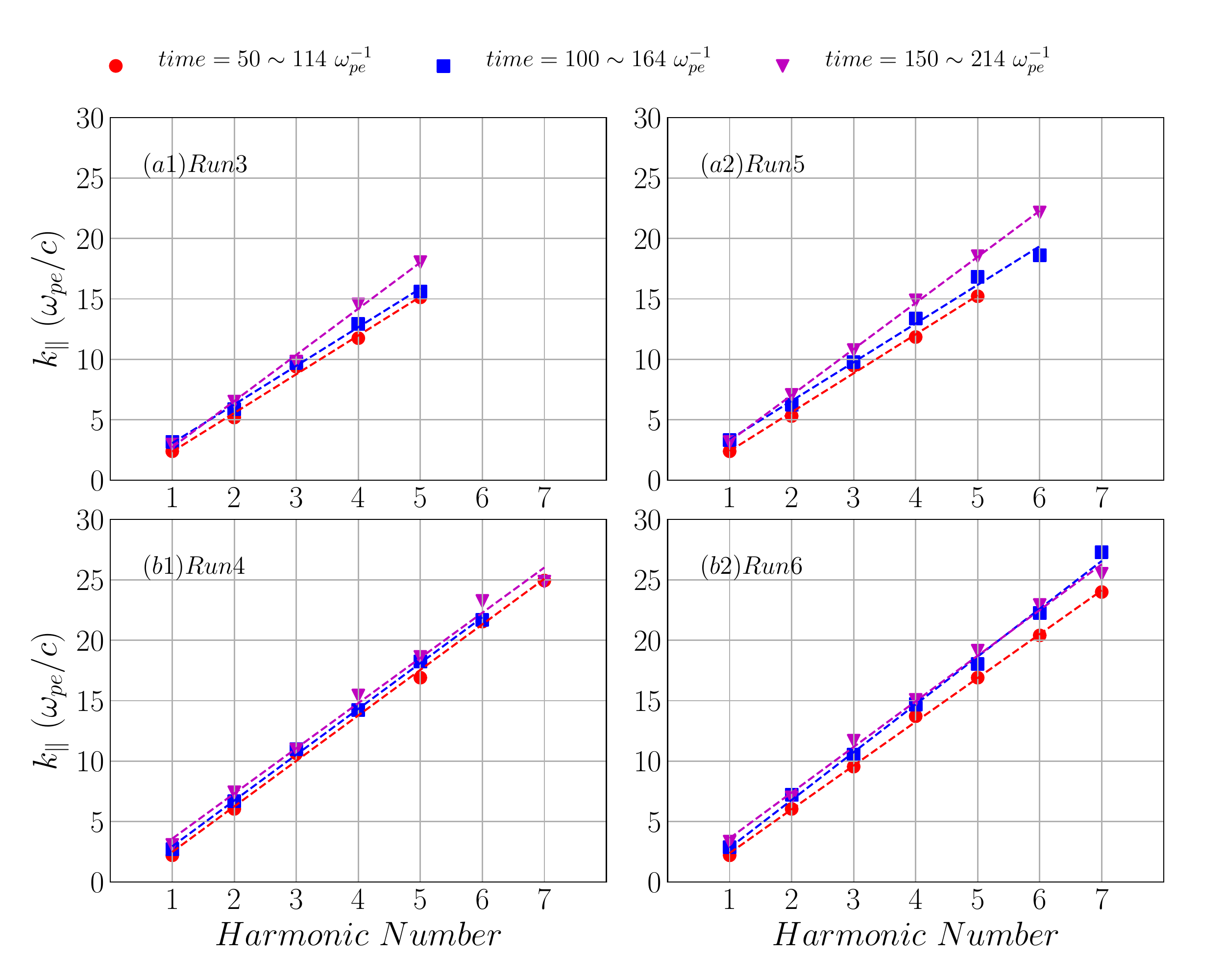}
    \caption[]{Linear regression analysis of characteristic wavenumbers of harmonics of Langmuir waves for Runs 3, 4, 5 and 6 in three consecutive time periods $50-114\ \omega_{pe}^{-1}$ (red dots), $100-164\ \omega_{pe}^{-1}$ (blue squares) and $150-214\ \omega_{pe}^{-1}$ (magenta triangles) respectively. The horizontal axis denotes the harmonic number $n$.}
    \label{fig:Wavevector_Temporal_Regression}
\end{figure}

Linear regression analysis of the characteristic wavenumbers $k_n$ of Langmuir modes is carried out at different time stages. \myreffig{fig:Wavevector_Temporal_Regression} shows the results of Runs 3,\ 4,\ 5,\ 6 during three consecutive time periods $t = 50-114,\ 100-164,\ 150-214\ \omega_{pe}^{-1}$.
We find that not only the characteristic wavenumber of (fundamental) Langmuir mode increases as the harmonic number increases, but also that of each Langmuir harmonic slightly increases over time. The latter is already confirmed in \myreffig{fig:PSD_El_Temporal}. It always agrees very well with the beat conditions of the wavenumber $k_n=nk_1$ and thus supports \myrefeq{eq:new_mechanism_LHs}.

\begin{figure}
    \centering
    \includegraphics[width=0.85\textwidth]{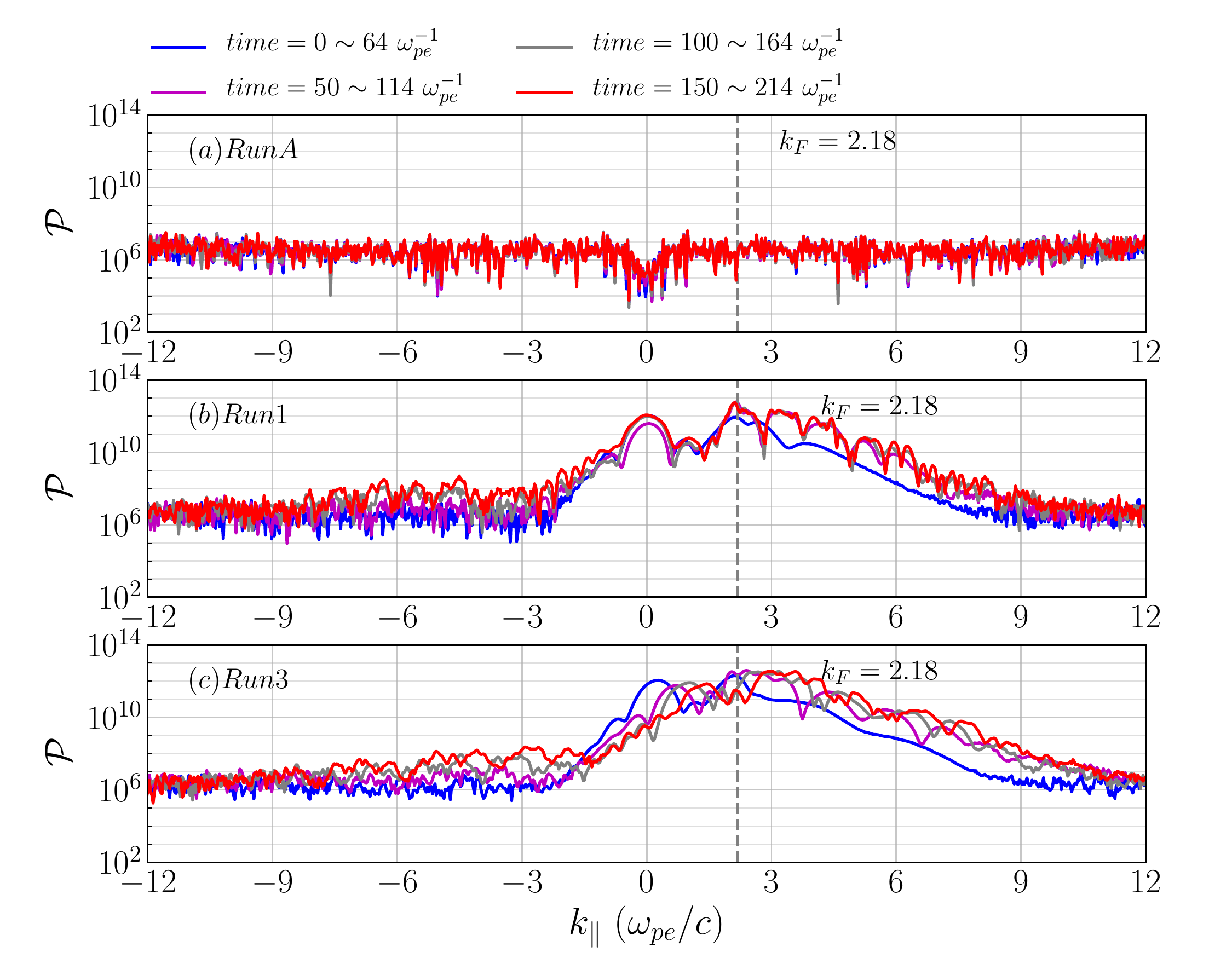}
    \caption[]{Power distribution $\mathcal{P}(k_{\parallel})$ (in unit of $statV^2\cdot cm^{-1}$) of fundamental Langmuir mode for Runs A, 1 and 3 in different time windows $t=0-64,\ 50-114,\ 100-164,\ 150-214\ \omega_{pe}^{-1}$, respectively. Here $k_F=2.18\ \omega_{pe}/c$ (denoted by the vertical dashed line) is the characteristic wavenumber of the beam-generated Langmuir mode.}
    \label{fig:PSD_FHZ_Temporal}
\end{figure}

We also find evidence of waves near $k_{\parallel}\to0$ in our simulations, i.e., for long-wavelengths.
Waves with $k_{\parallel}=0$ cannot of course take place in a numerical simulation with a finite-size simulation box, but it is nonetheless interesting to analyze the behavior of the spectral power for those small $k$ wavenumbers. 
\myreffig{fig:PSD_FHZ_Temporal} shows the power distribution of the fundamental Langmuir wave mode calculated by \myrefeq{eq:power_Langmuir} for RunA, Run1 and Run3 at different time windows.
Here RunA was carried out as a convergence test, where no electron beam is present (for details see Appendix \ref{app:CT}). 
The power $\mathcal{P}(k_{\parallel})$ of the fundamental Langmuir mode for long wavelengths ($k_{\parallel}\to 0$) of Run1 (with beam, $\eta=1$) is several orders of magnitude larger than that of RunA (without beam), which is comparable to the noise level (see \myreffig{fig:PSD_FHZ_Temporal}(a)).
The power of the fundamental Langmuir mode of Run1 is consistently larger than the noise level of RunA for all time intervals.
It shows a well defined local maximum in the long-wavelength limit (see \myreffig{fig:PSD_FHZ_Temporal}(b)). 
However, Run3 (with beam, $\eta=5$) has a clearly defined local maximum near $k_{\parallel}\to 0$ only during the initial time period (i.e., $0-64\ \omega_{pe}^{-1}$).
Indeed, \myreffig{fig:PSD_FHZ_Temporal}(c) shows that density inhomogeneities of background plasma inhibit these long wavelengths waves.
The wavevector $k_F$ in \myreffig{fig:PSD_FHZ_Temporal} is the characteristic wavenumber of beam-generated Langmuir mode, being determined by the location where the dispersion relation curve of the beam mode and of the fundamental Langmuir mode meet. In our simulations, $k_F\approx 2.18\ \omega_{pe}/c$. The power distribution $\mathcal{P}(k_{\parallel})$ at $k_F$ is similar for both Run1 and Run3. 
The peaks of these power enhancements move slightly to larger $k_{\parallel}$ over time. 

The enhancement of spectral power in that long-wavelength region may be due to waves with a wavenumber equal or smaller (i.e.: unresolved) than the spectral resolution, which is $k_{\parallel}=2\pi/L_x=0.03\ d_e^{-1}$. Another source for waves in this region can be due to the non-zero initial net current carried by the electron beam, causing a standing wave oscillating at the plasma frequency (i.e., plasma oscillations).
This is in turn a direct consequence of the displacement current in the Ampere's law  and the periodic boundary conditions.
Although it could be possible to suppress this initial net current by subtracting it at each timestep of the simulation, the physical consequences of such a procedure are even more unclear in our plasma system with a localized beam.

\subsection{Electron cyclotron waves}\label{sec:electroncyclotronwaves}

\begin{figure}
    \centering
    \includegraphics[width=0.85\textwidth]{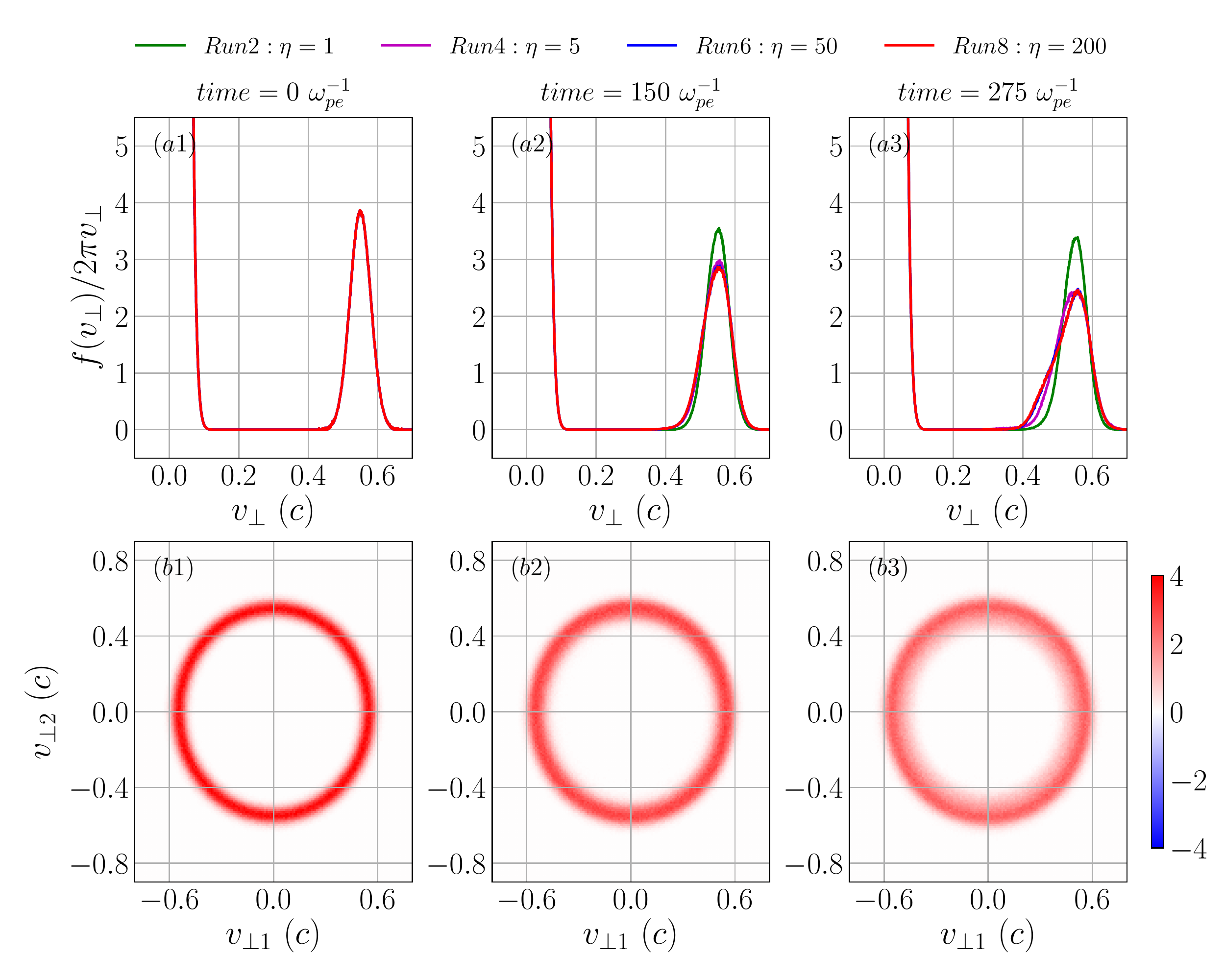}
    \caption[]{(a1-a3) Temporal evolution of perpendicular EVDF for Runs 2, 4, 6, 8 (ring beam) at time: $t=0,\ 150,\ 275\ \omega_{pe}^{-1}$ respectively. (b1-b3) Normalized 2D beam EVDF in the $v_{\perp1}\ vs.\ v_{\perp2}$ plane for Run6 at the same three moments of time.}
    \label{fig:perpprofile}
\end{figure}

The EVDF of a ring-beam can offer a perpendicular source of free energy  (positive gradient in the perpendicular EVDF).
This can cause ECM instabilities and generate electron cyclotron waves via the ECME mechanism. Simulations Runs 2, 4, 6, 8 address the non-linear evolution of localized ring-beam propagating in background plasmas with varying density gradients $\eta$.
\myreffig{fig:perpprofile} depicts perpendicular EVDF for Runs 2, 4, 6, 8 at three different moments of time $t=0,\ 150$ and $275\ \omega_{pe}^{-1}$ respectively (see \myreffig{fig:perpprofile}(a1-a3)). Results show that as time evolves the electron beams relax, i.e., the positive gradients in these perpendicular EVDFs are reduced, with the consequent transfer of its kinetic energy to plasma heating and wave energy. This relaxation occurs faster as the background density gradient becomes steeper (e.g., compare green curves for $\eta=1$ with the red curves for $\eta=200$), indicating a more efficient energy transfer. \myreffig{fig:perpprofile}(b1,b2,b3) show the time evolution of the 2D EVDF in $v_{\perp1}\ vs.\ v_{\perp2}$ plane for Run6 ($\eta = 50$). The 2D EVDFs clearly show the characteristic broadening of the ring beam over time, although still far from its complete relaxation. A considerable part of the free energy of the ring beams is transferred to growing plasma waves.

\begin{figure}
    \centering
    \includegraphics[width=0.85\textwidth]{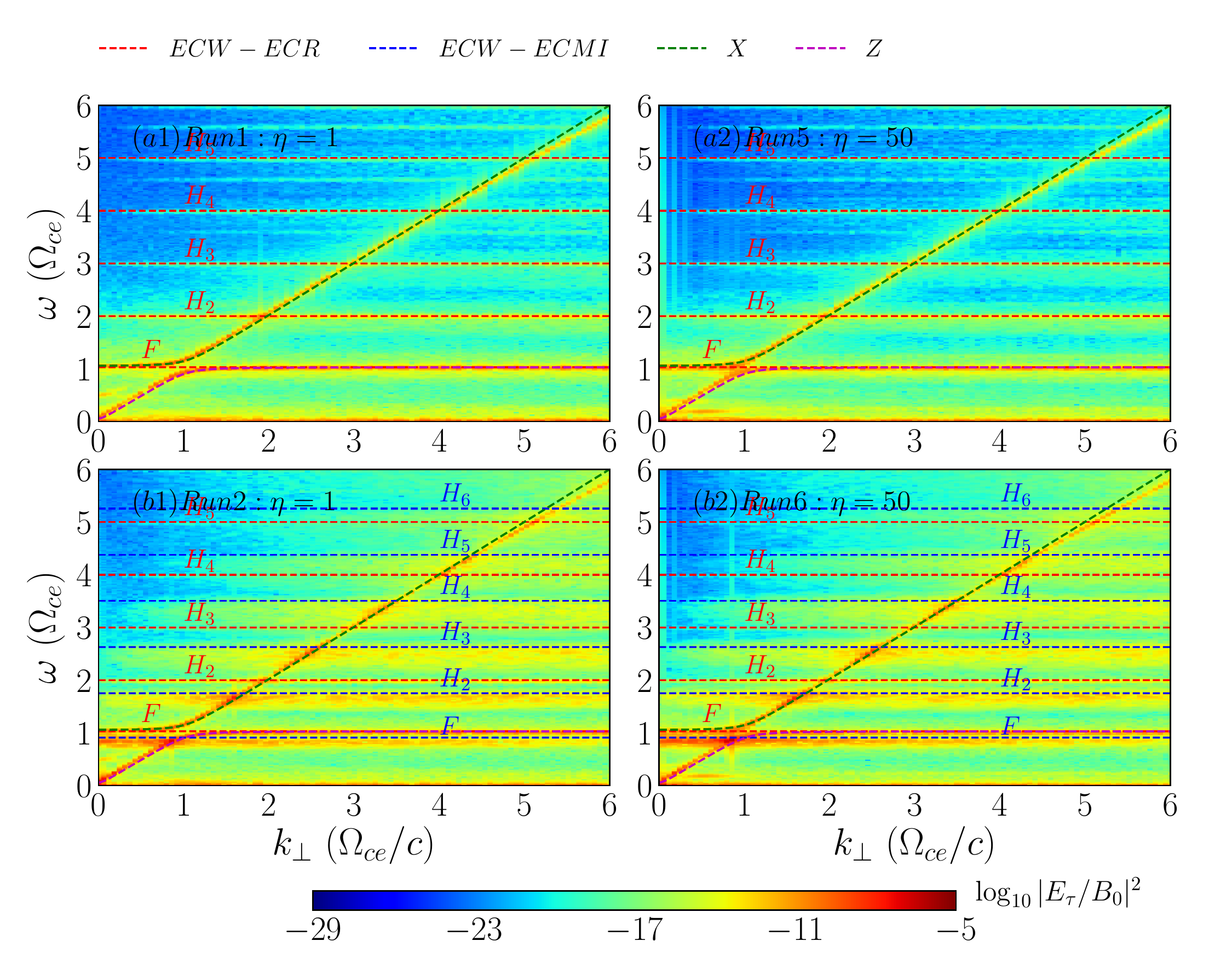}
    \caption[]{Illustration of fundamental (F) and harmonics (H) of electron cyclotron wave (ECW) modes. PSD in the $k_{\perp}\ vs.\ \omega$ plane for Runs 1,\ 2,\ 5,\ 6 in the time window $t=100-164\ \omega_{pe}^{-1}$ are displayed. The PSD is evaluated by $\log_{10}|E_{\tau}/B_0|^2$. Dispersion relation curves of fundamental and harmonics of electron cyclotron waves due to electron cyclotron resonances (ECRs, denoted by red dashed lines) and electron cyclotron maser instabilities (ECMIs, denoted by blue dashed lines), X (green dashed curve) and Z (magenta dashed curve) modes are overlaid.}
    \label{fig:CyclotronHF}
\end{figure}

\myreffig{fig:CyclotronHF} depicts the PSD of the transverse electric field in ($\omega, k_{\perp}$) domain during $t=100-164\ \omega_{pe}^{-1}$ for simulations Runs 1 and 2 ($\eta=1$), as well as 5 and 6 ($\eta=200$). Note that Runs 1 and 5 are carried out with a Maxwellian beam, while Runs 2 and 6 with a ring beam, so that we can compare the effect perpendicular sources of free energy (positive gradients in the EVDFs).
\myreffig{fig:CyclotronHF} shows that fundamental and harmonics of electron cyclotron waves due to electron cyclotron resonances (ECRs) are observed in all the simulations (e.g., Runs 1, 2, 5, 6, see red dashed lines in \myreffig{fig:CyclotronHF}).
On the other hand, fundamental and harmonics of electron cyclotron waves due to ECM instabilities are only observed in simulations with a ring beam (e.g., Runs 2, 6, see blue dashed lines in \myreffig{fig:CyclotronHF}(b1,b2)). The dispersion relations of various electron cyclotron waves will be discussed below. Results show that those waves, including high-order harmonics, are already generated at the early stage of the instability growth (e.g., $t=100-164\ \omega_{pe}^{-1}$) and well before the relaxation of the velocity gradient in perpendicular EVDF starts (see \myreffig{fig:perpprofile}).
By comparing results between Runs 1 and 5 and between Runs 2 and 6, we notice that the background density gradients practically have no influence on the generation of electron cyclotron waves.

As mentioned above, two mechanisms contribute to the generation of the fundamental and harmonics of electron cyclotron waves. The first one is due to the electron cyclotron resonances (ECRs) at $\omega=n\Omega_{ce}\ (n=1,2,3\dots)$ (see red dashed lines in \myreffig{fig:CyclotronHF}) and the other is attributed to the electron cyclotron maser instabilities (ECMIs) caused by the ring beam (see blue dashed lines in \myreffig{fig:CyclotronHF}(b1,b2)). We discuss now their corresponding dispersion relations.

The dispersion relation of the well known electron cyclotron Bernstein modes is expressed in the following form \citep[see, e.g.,][]{Melrose1986}

\begin{equation}
    \begin{aligned}
        1-\frac{2\omega_{pe}^2}{\lambda_e}e^{-\lambda_e}\sum\limits_{n=1}^{\infty}\frac{n^2I_n(\lambda_e)}{\omega^2-n^2\Omega_{ce}^2}=0\label{eq:ElectronBernsteinMode_ECR}
    \end{aligned}
\end{equation}
where $I_n(\lambda)$ is the modified Bessel function of the first kind with argument $\lambda_e=k^2v_{the}^2/\Omega_{ce}^2$, and $v_{the}$ is the electron thermal speed of the background plasma. In the limit $\lambda_e\ll 1$, $e^{-\lambda_e}\rightarrow 1$ and the asymptotic expression of the modified Bessel function $I_n(\lambda)$ becomes $I_n(\lambda_e)\approx \frac{1}{n!}\left(\frac{\lambda_e}{2}\right)^n$. Using this approximation of $I_n(\lambda_e)$ in \myrefeq{eq:ElectronBernsteinMode_ECR} we can get approximate solutions for the frequency $\omega\approx\sqrt{\omega_{pe}^2+\Omega_{ce}^2}$, $2\Omega_{ce}$, $3\Omega_{ce},\dots$. While in the opposite limit of $\lambda_e\gg 1$, the modified Bessel function $I_n(\lambda)$ can be approximated as $e^{-\lambda_e}I_n(\lambda_e)\approx \frac{1}{\sqrt{2\pi\lambda_e}}e^{-n^2/2\lambda_e}$. By solving \myrefeq{eq:ElectronBernsteinMode_ECR} for the frequency with this approximated expression, we get solutions for the harmonics of the electron cyclotron frequency as $\omega=n\Omega_{ce}$, $n=1,2,3,\dots$.

The dispersion relation of electron cyclotron wave modes by ECMIs is obtained by a small modification related to relativistic effects in~\myrefeq{eq:ElectronBernsteinMode_ECR}, yields
\begin{equation}
    \begin{aligned}
        1-\frac{2\omega_{pe}^2}{\lambda_e^2\gamma_{\perp}^2}e^{-\lambda_e^2\gamma_{\perp}^2}\sum\limits_{n=1}^{\infty}\frac{n^2I_n(\lambda_e^2\gamma_{\perp}^2)}{\omega^2-n^2\Omega_{ce}^2/\gamma_{\perp}^2}=0\label{eq:ElectronBernsteinMode_ECME}
    \end{aligned}
\end{equation}
The reason is that those electron cyclotron wave modes are associated to velocity gradient in the perpendicular EVDF of ring beam. Relativistic effects related to the perpendicular drift speed of the beam are essential for this maser mechanism \citep{Melrose2017a}.
Specifically, the difference between the standard \myrefeq{eq:ElectronBernsteinMode_ECR} and their ECM counterpart \myrefeq{eq:ElectronBernsteinMode_ECME} has to do with the the perpendicular Lorentz factor $\gamma_{\perp}$ derived from the perpendicular drift speed of the ring beam, i.e., 
\begin{align}
    \gamma_{\perp}=\sqrt{1+u_{d\perp}^2/c^2}\label{eq:perpLorentzfactor}
\end{align}
Accordingly, the electron cyclotron frequency term $\Omega_{ce}$ in \myrefeq{eq:ElectronBernsteinMode_ECR} has to be replaced by $\Omega_{ce}/\gamma_{\perp}$, consequently, to get \myrefeq{eq:ElectronBernsteinMode_ECME}.
The consequence of this difference is a downward shift of the resonance frequencies, i.e., the ECMI-related dispersion curves (see blue dashed lines in \myreffig{fig:CyclotronHF} (b1,b2)) are located always below the ECR-related dispersion curves (see red dashed lines in \myreffig{fig:CyclotronHF} (b1,b2)).

\begin{figure}
    \centering
    \includegraphics[width=0.85\textwidth]{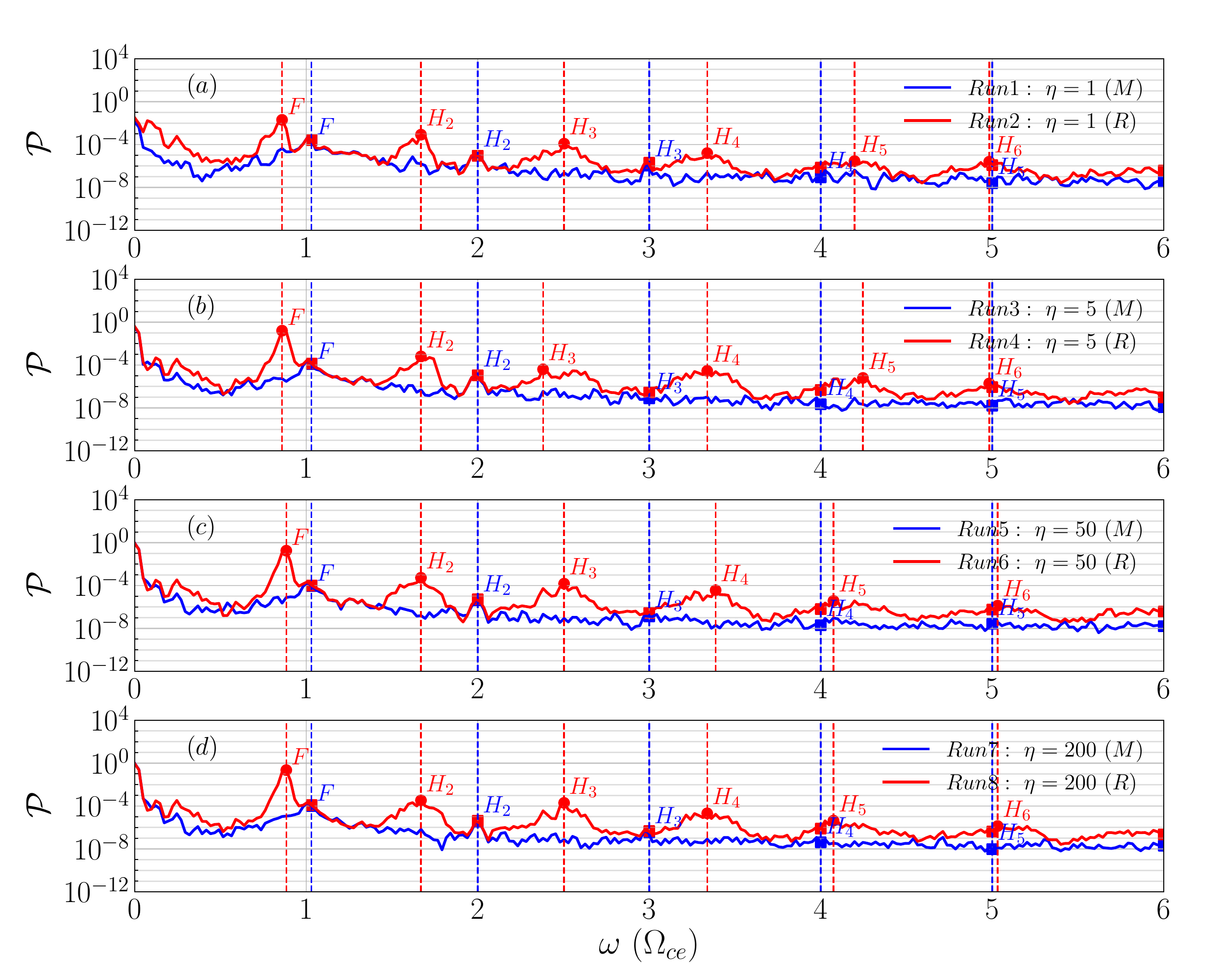}
    \caption[]{Power spectrum derived from transverse electric field $E_{\tau}$ for all simulations during $t=100-164\ \omega_{pe}^{-1}$. The coloured squares denote harmonics of electron cyclotron mode due to ECRs and coloured dots indicate those due to ECM instabilities. Other labels and notations are used in the same way as in \myreffig{fig:Energy_LangmuirwavesHF}.}
    \label{fig:Energy_Cyclotron}
\end{figure}

Note that a numerical aliasing effect appears in the form of line-like PSD enhancements in \myreffig{fig:CyclotronHF} (a1,a2) between frequencies $\omega=3$ and $4\ \Omega_{ce}$, $4$ and $5\ \Omega_{ce}$ as well as between $5$ and $6\ \Omega_{ce}$.
This is an artefact due to our maximum sampling frequency of $2\pi\Omega_{ce}$. Hence, for any wave mode with wave frequency above this frequency, i.e., $\omega>2\pi\Omega_{ce}\approx 6 \Omega_{ce}$, a folding to lower frequency region (i.e., in which $\omega<2\pi\Omega_{ce}$) takes place. We also notice that in the large wavevector regime (e.g., $k_{\perp}>4.5\ \omega_{ce}/c$) the analytical dispersion relation curve of X mode deviates corresponding enhancement in PSD slightly for each simulation (see \myreffig{fig:CyclotronHF}). The deviations between analytical dispersion relation curves and simulation results, in particular in the larger wavevector and frequency regime, are inherent in numerical simulations due to the usage of finite grid cell size and finite time step.

To compare the characteristic frequencies of electron cyclotron wave modes due to the different mechanisms mentioned above, we calculated the power spectrum by integrating the PSD of the transverse electric field over perpendicular wavevector $k_{\perp}$ in $(\omega,k_{\perp})$ domain, i.e.,
\begin{equation}
    \mathcal{P}_{\tau}(\omega)=\int |E_{\tau}\left(\omega,k_{\parallel}=0,k_{\perp}\right)|^2dk_{\perp}\label{eq:power_frequency}
\end{equation}

\myreffig{fig:Energy_Cyclotron} shows the resulting power spectra  for all eight simulations in the time window $100-164\ \omega_{pe}^{-1}$. The numbered peaks indicate the harmonics of electron cyclotron waves mode. The blue/red squares correspond to harmonics of electron cyclotron waves due to ECRs while red dots indicate harmonics due to the ECM instabilities.
For electron cyclotron waves caused by ECRs, the power spectra generated by Maxwellian and by ring beam are practically the same at the characteristic frequencies of these wave modes.
While for electron cyclotron waves produced by ECM instabilities, the power spectra generated by the ring beam significantly exceeds that generated by the Maxwellian beam at their respective frequencies.
\myreffig{fig:Energy_Cyclotron} further confirms that the density inhomogeneities of the background plasma have no influence on the generation of electron cyclotron waves.

\subsection{Doppler frequency shift}

\begin{figure}
    \centering
    \includegraphics[width=0.85\textwidth]{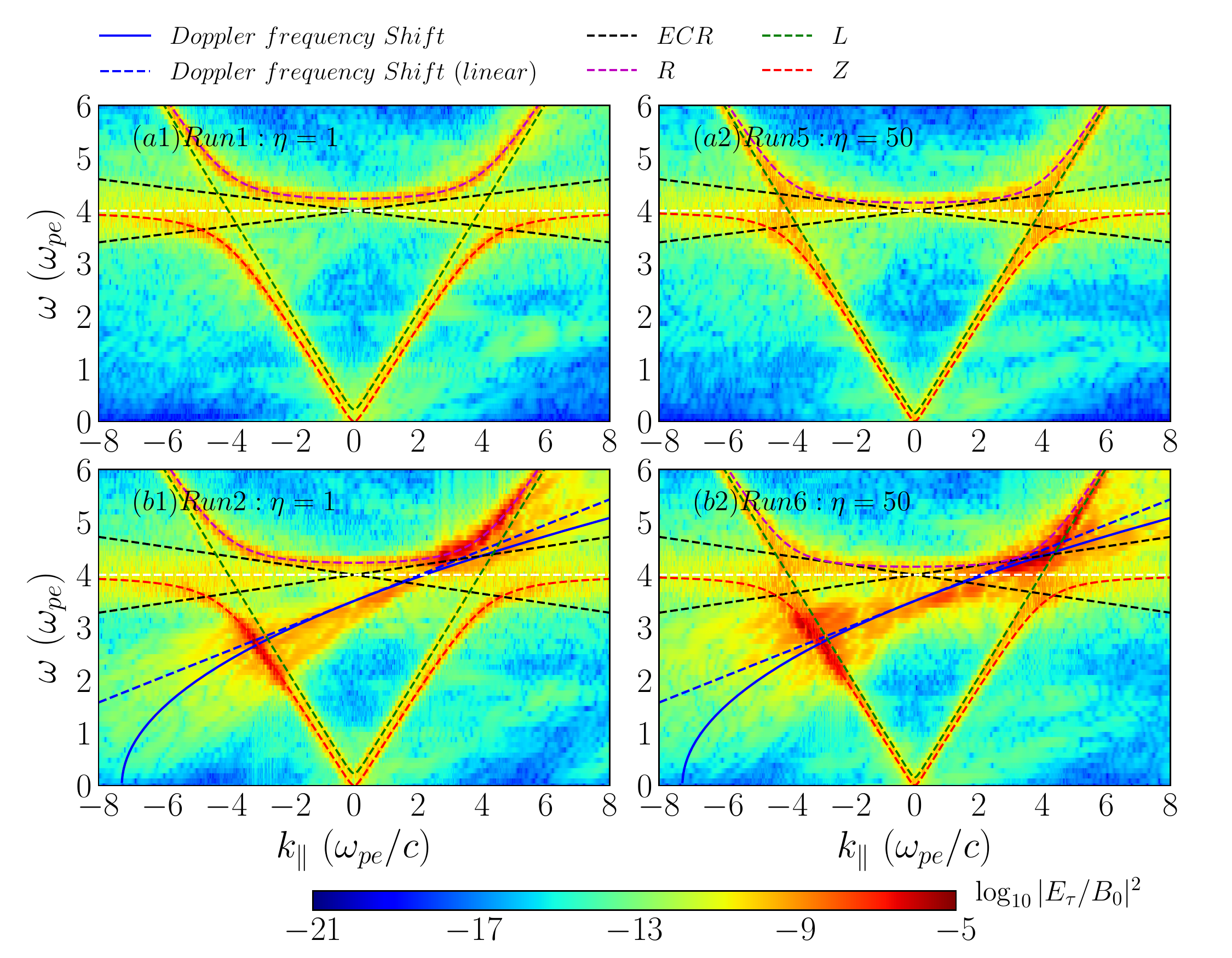}
    \caption[]{PSD in $k_{\parallel}\ vs.\ \omega$ plane for Runs 1,\ 2,\ 5,\ and 6 in the time interval $t=100-164\ \omega_{pe}^{-1}$. The PSD is evaluated as $\log_{10}|E_{\tau}/B_0|^2$. The dispersion relation curves of the R (magenta dashed curve), L (green dashed curve) and Z (red dashed curve) modes as well as those predicted by \myrefeq{eq:Dopplershift} (blue solid curve) and \myrefeq{eq:Dopplershift_linear} (blue dashed line) are overlaid.}
    \label{fig:kpara_Et}
\end{figure}

\myreffig{fig:kpara_Et} shows the PSD of transverse electric field $E_{\tau}$ for the simulations with homogeneous background plasma (i.e., Runs 1 and 2, $\eta=1$, see \myreffig{fig:kpara_Et}(a1,b1)) as well as with inhomogeneous background plasma (i.e., Runs 5 and 6, $\eta=50$, see \myreffig{fig:kpara_Et}(a2,b2)). 
As shown in \myreffig{fig:kpara_Et}, electron thermal fluctuations develop near the electron cyclotron frequency inside triangular regions delimited by black dashed lines, which are given by 
\begin{align}\label{eq:resonances}
    \omega=\Omega_{ce}\pm 3v_{the}k_{\parallel}
\end{align}
For small $|k|\leq 5\ \omega_{pe}/c$ the Z mode (indicated by a red dashed curve in \myreffig{fig:kpara_Et}) does not enter the triangular regions. 
In the short wavelength regime $|k|>5\ \omega_{pe}/c$, the Z mode asymptotically approaches the electron cyclotron frequency $\Omega_{ce}=4\ \omega_{pe}$ and enters into the triangular regions, becoming heavily damped.

For the simulations with a ring-beam, Runs 2 and 6, a significant enhancement of power near the Doppler-shifted electron-cyclotron resonance region is observed (see \myreffig{fig:kpara_Et}(b1,b2)).
This region with enhanced power crosses the Z and L wave modes near frequencies $\omega\approx 2-3\ \omega_{pe}$ at $k_{\parallel}\approx -4\ \omega_{pe}/c$, and crosses the L and R wave modes above the ECR frequencies for $\omega\approx 4-6\ \omega_{pe}$ at $k_{\parallel}\approx 4\ \omega_{pe}/c$. 
These PSD enhancements can be fitted by an empirical expression  (see blue solid curves denoted by ``Doppler frequency Shift'' in \myreffig{fig:kpara_Et}(b1,b2)):
\begin{align}
    \omega=\frac{1}{\gamma_{\perp}}\sqrt{\Omega_{ce}\left(\Omega_{ce}+u_{d\perp}k_{\parallel}\right)}\label{eq:Dopplershift}
\end{align}
In the long wavelength limit, i.e. for $k_{\parallel}\to 0$, the fitting \myrefeq{eq:Dopplershift} reveals another linear expression (see blue dashed oblique lines denoted by ``Doppler frequency shift (linear)'' in \myreffig{fig:kpara_Et}(b1,b2)):
\begin{align}
    \omega\approx \frac{1}{\gamma_{\perp}}\left(\Omega_{ce}+\frac{1}{2}u_{d\perp}k_{\parallel}\right)\label{eq:Dopplershift_linear}
\end{align} 

The fittings \myrefeq{eq:Dopplershift} and \myrefeq{eq:Dopplershift_linear} indicate that the power enhancements are attributed to a Doppler frequency shift of waves at the electron cyclotron frequency associated to the beam drift speed.

\begin{figure}
    \centering
    \includegraphics[width=0.85\textwidth]{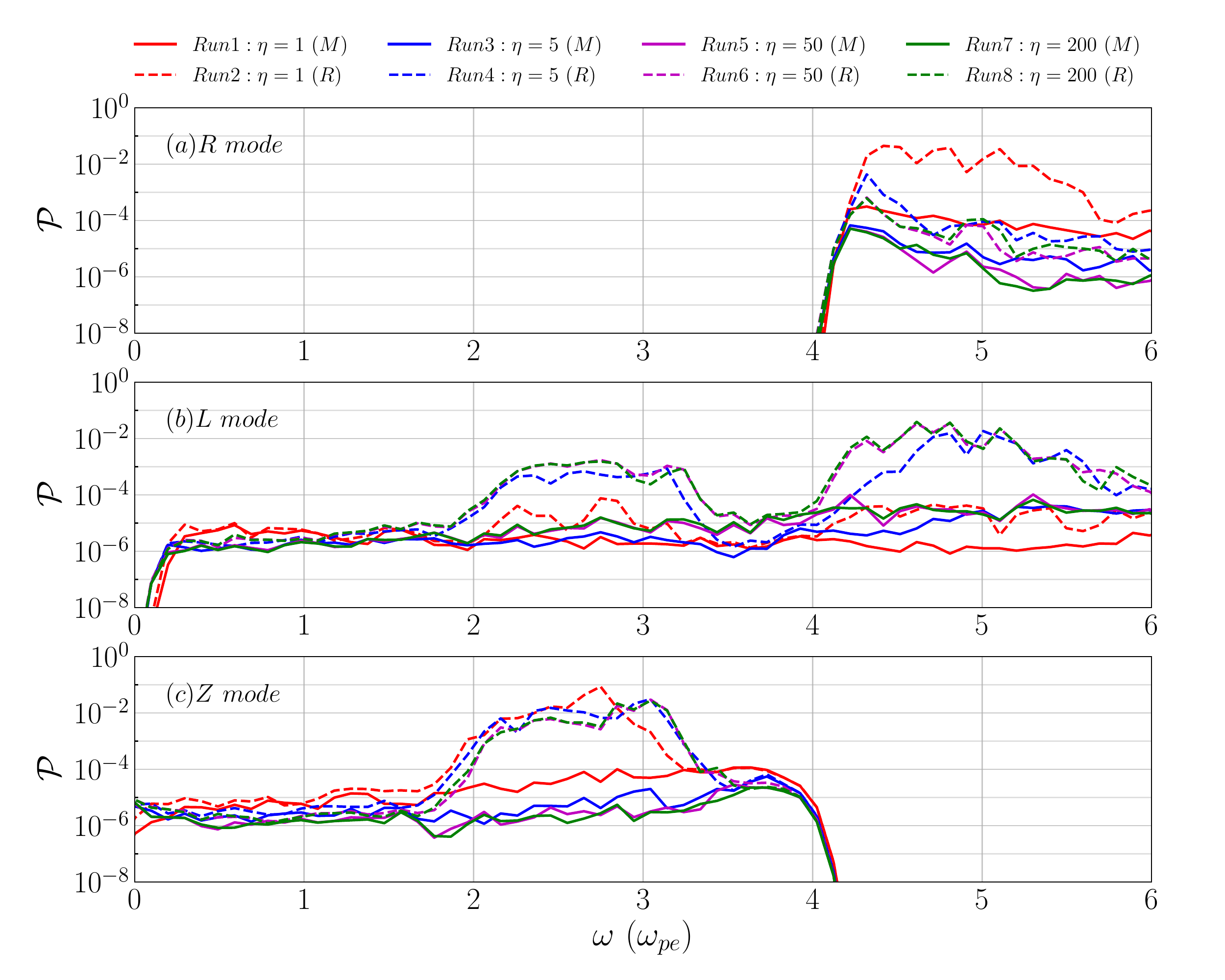}
    \caption[]{Power spectrum (in unit of $statV^2\cdot cm^{-2}\cdot s$) of R (a), L (b) and Z (c) modes for all simulations in $t=100-164\ \omega_{pe}^{-1}$.}
    \label{fig:Energy_Et}
\end{figure}

\myreffig{fig:Energy_Et} shows the power spectrum of the R (see \myreffig{fig:Energy_Et}(a)), L (see \myreffig{fig:Energy_Et}(b)) and Z (see \myreffig{fig:Energy_Et}(c)) wave mode respectively. These power spectra are calculated from the transverse electric field according to \myrefeq{eq:power_frequency} in the time window $t=100-164\ \omega_{pe}^{-1}$. 
Results show that, for the ring beam cases (i.e., Runs 2, 4, 6, 8, see dashed curves in \myreffig{fig:Energy_Et}), the power of the R mode wave is enhanced mainly in the frequency range  $\omega=4-6\ \Omega_{ce}$, while the L mode is enhanced for both $2-3\ \Omega_{ce}$ and $4-6\ \Omega_{ce}$. The Z wave mode is enhanced mainly for $2-3\ \Omega_{ce}$. 
The density inhomogeneities also influence the power spectra of wave modes. With the increase of the density gradient $\eta$, the power of the R mode decreases from about $10^{-1}$ to $10^{-6}$ in the frequency range $\omega=4-6\ \Omega_{ce}$. The same happens to the power of the Z mode: its power decreases from about $10^{-1}$ to $10^{-2}$ in the frequency range $\omega=2-3\ \Omega_{ce}$ as the density gradient increases.
The L mode has the opposite behavior: its power increases from about $10^{-6}$ to $10^{-3}$ for $\omega=2-3\ \Omega_{ce}$ and from about $10^{-6}$ to $10^{-1}$ in the range $\omega=4-6\ \Omega_{ce}$ as the density gradient increases.

\section{Conclusions}\label{sec:conclusions}

By utilizing fully-kinetic PIC-code simulations, we investigated the influence of perpendicular and parallel gradients in the EVDFs of localized electron beams. 
Such distribution functions, as well as density inhomogeneities (cavities), are presumably formed by magnetic reconnection in solar flares. The influence of density inhomogeneities on the wave generation is  investigated. 
The electron beams under investigation generate unstable plasma waves via wave-wave and wave-particle (electron cyclotron maser) interactions. Those plasma waves could lead to electromagnetic radio emission.

We found the following effects of density inhomogeneities along the path of propagating electron beams on radio emissions:

(1) In density cavities with low density and strong magnetic field, high-order harmonic Langmuir waves are generated by bump-on-tail EVDF parallel to the magnetic field direction in velocity space. 
The stronger the density gradients are, the higher harmonics are generated. To the best of our knowledge this effect has not been observed before. 
In combination with model of electron transport and wave propagation \citep[see, e.g.,][]{Li2009}, this effect of density inhomogeneities would be able to explain the observed harmonics of Langmuir waves in solar radio bursts.

(2) The stronger the density gradients are, the more efficient is the energy conversion from the beam EVDF into plasma heat and energy of electromagnetic waves. The free energy is related to the parallel and perpendicular gradients in the beam EVDF.

(3) Density gradients along the propagation path of electron beams cause a downward shift of the characteristic frequencies of the Langmuir waves and their harmonics. 
This effect is attributed to the variation of the local plasma frequency along the path of the electron beam through the inhomogeneous ambient plasma.

(4) Perpendicular velocity gradients in the beam EVDF cause a significant Doppler frequency shift near the electron cyclotron resonance region due to relativistic effects.

The generation of waves at the harmonics of the plasma frequency is associated with wave-wave coupling referred to as the plasma emission mechanism. The generation of waves at harmonics of the electron cyclotron frequency is associated with electron cyclotron resonances and with the electron cyclotron maser mechanism.

Based on our results, the generation of harmonics of Langmuir waves can be understood in the following way: any $n$-th harmonic Langmuir wave mode $L_n$ is generated by an interaction of adjacent harmonic mode $L_{n-1}$ with the beam-generated (fundamental) Langmuir wave mode $L$, i.e., $L_{n-1}+L\to L_n$. We obtained a semi-empirical expression for the dispersion relation of higher harmonic of Langmuir waves generated in density inhomogeneities along the path of the beam propagation. 

Note that our results deal only with the initial stage of wave generation and beam propagation at electron time-scales. During this initial interval the wave-wave interaction depending on ion acoustic ($S$) waves does not start yet, since those waves develop only at longer (ion) time-scales. 
That is incompatible with the computational restrictions due to our chosen parameters,
since such a simulation would require a simulation box much larger.
This restriction of our results is due to the goal of investigating the role of inhomogeneities through which mildly-relativistic and spatially-localized beams quickly propagate. 
For a direct comparison with observations, the initial wave generation mechanism studied here needs to be combined with a global (large-scale) model of wave propagation and electron transport in the solar corona and in the solar wind ~\citep[see, e.g,][]{Li2009,Reid2018}. 
Our findings are, nevertheless, relevant for understanding the kinetic physics of the generation of plasma waves by pure bump-on-tail and ring-beam EVDFs in inhomogeneous plasmas.

The other caveat of our studies, as well as most of other studies dealing with the ECM instabilities, is the problem of explaining how these radio waves escape from the low-density regions (density cavities), where the condition $\Omega_{ce}>\omega_{pe}$ holds.
In our model, those regions are assumed to be along the the separatrices of guide-field magnetic reconnection.
Outside the cavities the plasma density is higher and the frequency ratio  $\Omega_{ce}/\omega_{pe}$ turns into the opposite ($<1$), which inhibits the further propagation of electromagnetic waves trapped inside the cavities.
There are several possible solutions to this escape problem of radio waves. This could be appropriate mode conversion processes \citep{Treumann2017} or the existence of solar coronal flux tubes. According to them, the cutoff frequency (i.e., the local plasma frequency) outside the flux tube equals the wave frequency, and the radio waves therefore can escape~\citep{Wu2002,Wu2014}.

\section{Acknowledgements}

We gratefully acknowledge the developers of the ACRONYM code, the \textit{Verein zur F\"orderung kinetischer Plasmasimulationen e.V.} and the financial support by the German Science Foundation (DFG), projects MU-4255/1-1 and BU 777/15-1.
We also acknowledge the Sino-German collaboration made possible thanks to the project with NSFC grant 11761131007.
We also gratefully acknowledge the possibility of using the computing resources of the Max Planck Computing and Data Facility (MPCDF, formerly known as RZG) at Garching and of the Max-Planck-Institute for Solar System Research at G\"ottingen as well as of the Technical University Berlin, Germany.
We also thank the referees for their comments and suggestions that allowed us to improve the presentation of our results.

\appendix
\par

\section{Results of Convergence Tests} 
\label{app:CT}

We verified that our model and the chosen numerical parameters are able to reproduce the physical processes described in this work without introducing numerical artifacts.
For this sake we varied the grid resolution and the number of macro-particles per cell.

The numerical setup of the convergence test simulations can be summarized in comparison with Run1 as follows:

\begin{itemize}
\item RunA, without the beam, only the background plasma, i.e., $N_{bg}=950$ (number of background macro-particles per cell) and $N_{bm}=0$  (number of beam macro-particles per cell).
\item RunB, with better grid resolution. It has the double of grid points along each direction, i.e., $N_x\times N_y = 8192\times 1024$ and the same physical size so that the grid cell size is half of that of Run1 (i.e., $\Delta x=\lambda_{D}$ instead of $\Delta x=2\lambda_{D}$). In addition, the number of macro-particles per cell is half of that of Run1, i.e., $N_{bg}=475,\ N_{bm}=25$.
\item RunC, with the double number of macro-particles per cell, i.e., $N_{bg}=1900,\ N_{bm}=100$.
\end{itemize}
See details of the differences between the setup of Run1 and these convergence tests in \myreftab{tab:CT_setup}.

\begin{center}
    \begin{threeparttable}
        \begin{tabular}{c|cc|cc|cc|r}
            \hline
            \multirow{2}{*}{Run} &\multicolumn{2}{|c}{macro-particles per cell} &\multicolumn{2}{|c|}{number of grid points} &\multicolumn{2}{|c|}{Physical size ($\lambda_{D}$)} & \multirow{2}{*}{$\Delta x$}\\
            \cline{2-7}
            &$N_{bg}$ & $N_{bm}$ & $N_x$ & $N_y$& $L_x$ & $L_y$ \\
            \hline
            1 & 950  & 50  & 4096 & 512 & 8192 & 1024  & 2$\lambda_{D}$\\
            A & 950  & 0   & 4096 & 512 & 8192 & 1024 & 2$\lambda_{D}$\\
            B & 475  & 25  & 8192 & 1024 & 8192 & 1024 & $\lambda_{D}$\\
            C & 1900 & 100 & 4096 & 512 & 8192 & 1024 & 2$\lambda_{D}$\\
            \hline
        \end{tabular}
        \caption{Input parameters of the additional simulations aiming at testing the numerical convergence. Here $N_{bg}$ and $N_{bm}$ are the number of macro-particles per cell of background and beam plasma, $N_x$ and $N_y$ are the number of grid cells of the 2D simulation box, and $\Delta x$ is the grid cell size.}
        \label{tab:CT_setup}
    \end{threeparttable}
\end{center}

\begin{figure}
    \centering
    \includegraphics[width=0.85\textwidth]{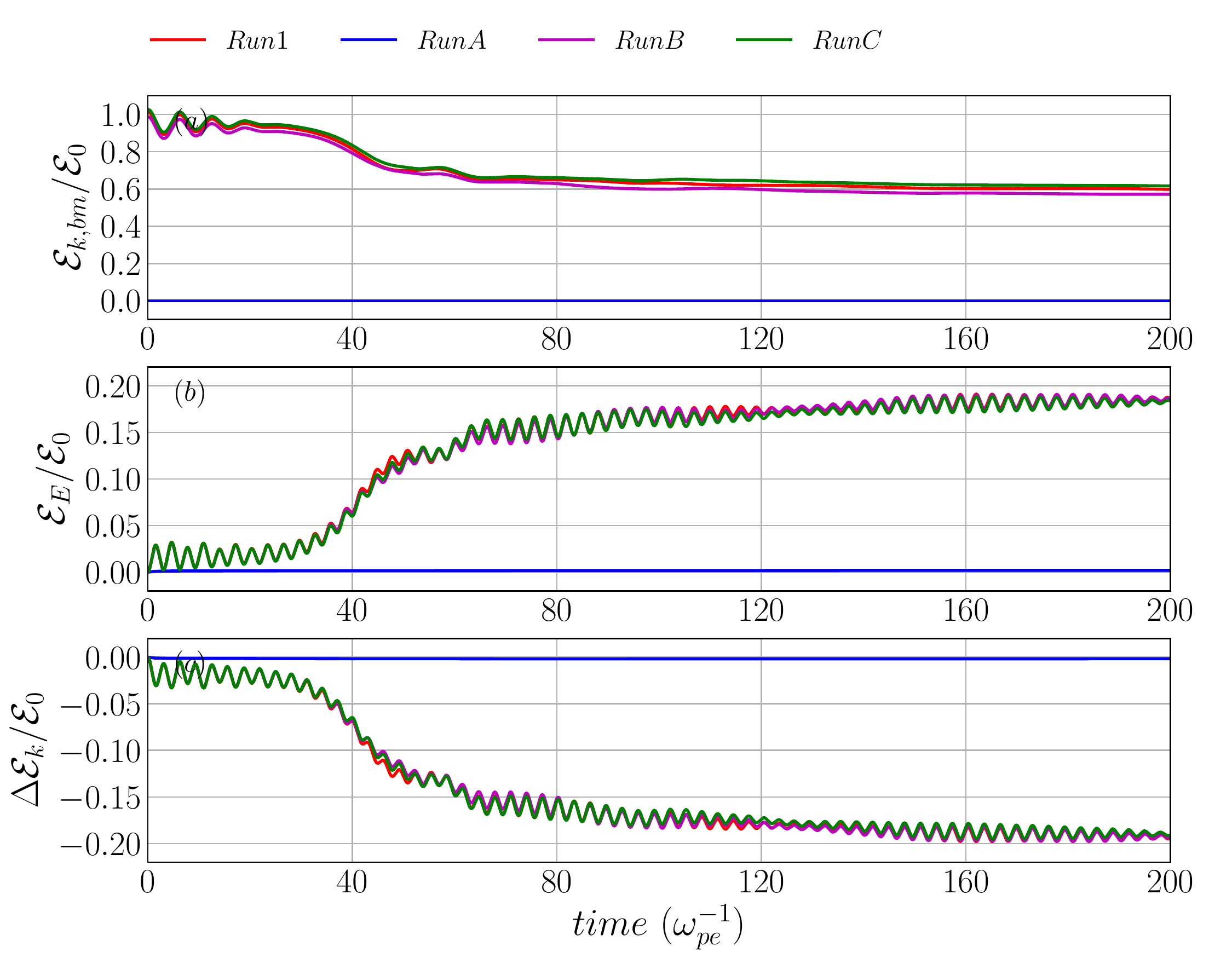}
    \caption[]{Comparison of the temporal evolution of different energy components for the different convergence tests RunA, RunB, RunC and the standard Run1. (a) Total kinetic energy of electron beam $\mathcal{E}_{k,bm}$, (b) electric field energy $\mathcal{E}_{E}$ and (c) energy variation of the total kinetic energy of the beam-plasma system $\mathcal{E}_{k}(t)-\mathcal{E}_{k,bm}(t=0)$ are displayed. All quantities are normalized by the initial total kinetic energy $\mathcal{E}_0=\mathcal{E}_{k,bm}(t=0)$ for Run1.}
    \label{fig:Ek_CT}
\end{figure}

ACRONYM is a standard momentum-conserving PIC code, so the energy is not conserved exactly, but to a very good extent.
In \myreffig{fig:Ek_CT} we estimated the difference in energy conservation by displaying the time evolution of the total kinetic energy, total beam kinetic energy and electric field energy of the standard Run1 and these new simulations RunA, RunB and RunC. The definitions of all those energies are given in Section~\ref{sec:conversion_energy}. Note that the total kinetic energy includes both bulk flow kinetic energy as well as thermal energy (i.e., according to the notation of Section~\ref{sec:conversion_energy}, $\mathcal{E}_t=\mathcal{E}_{th}+\mathcal{E}_b$).
The magnetic energy is not shown since their variations are negligible compared to all other energy components.
The total kinetic energy of the beam electrons $\mathcal{E}_{k,bm}$ is zero for RunA since there is no beam at all. 
\myreffig{fig:Ek_CT} shows that each individual energy component is consistent very well with those of the standard Run1 and with each other.
We can clearly see that, during the time period we concentrate on, i.e., $t=0-200\ \omega_{pe}^{-1}$, the loss of total kinetic energy of the beam-plasma system (mostly from the beam electrons) is mainly converted into electric energy.
The maximum relative change of the total energy $\Delta \mathcal{E}/\mathcal{E}_0$ (not shown here since it is very close to zero within the range of vertical axis) is about $0.07\%$ for the Run1 at the end of the considered time period. The new runs with a larger number of macro-particles per cell and/or grid resolution feature an even smaller change of the total energy: no more than half of that of Run1. So the total energy is relatively well conserved and not very sensitive to variation in those numerical parameters.

We also compare the effects of increasing the number of macro-particles per cell and spatial resolution on the power spectral density of electric fields fluctuations.

\begin{figure}
    \centering
    \includegraphics[width=0.85\textwidth]{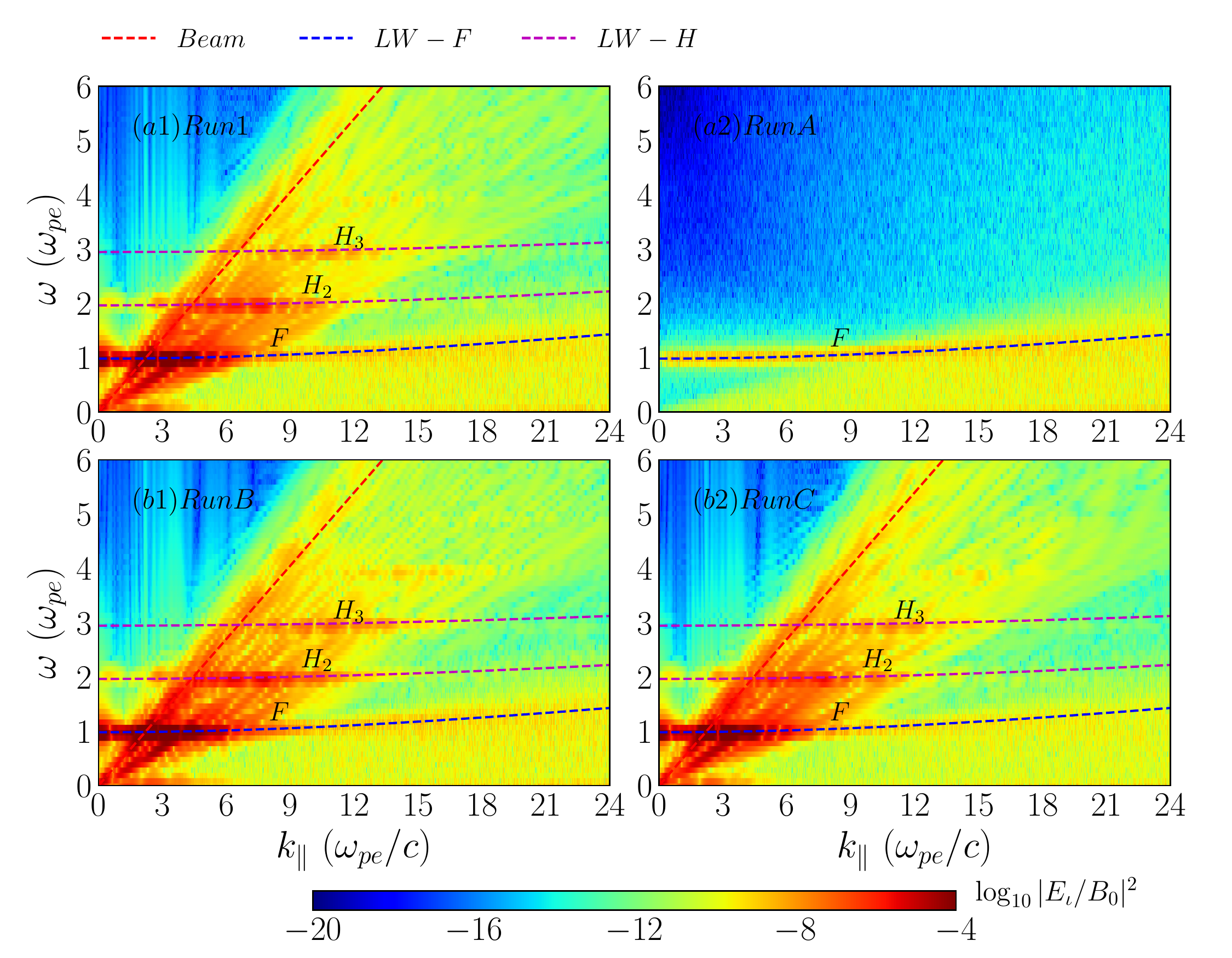}
    \caption[]{PSD in $k_{\parallel}\ vs.\ \omega$ plane for Runs 1, A, B and C in the time period $t=100-164\ \omega_{pe}^{-1}$. PSD is evaluated as $\log_{10}|E_{\iota}/B_0|^2$. Dispersion relation curves of the beam mode (red dashed line), fundamental (F, denoted by blue dashed curve) and harmonics (H, denoted by magenta dashed curves) of Langmuir wave (LW) modes are overlaid.}
    \label{fig:Energy_CT_El_kpara}
\end{figure}

\begin{figure}
    \centering
    \includegraphics[width=0.85\textwidth]{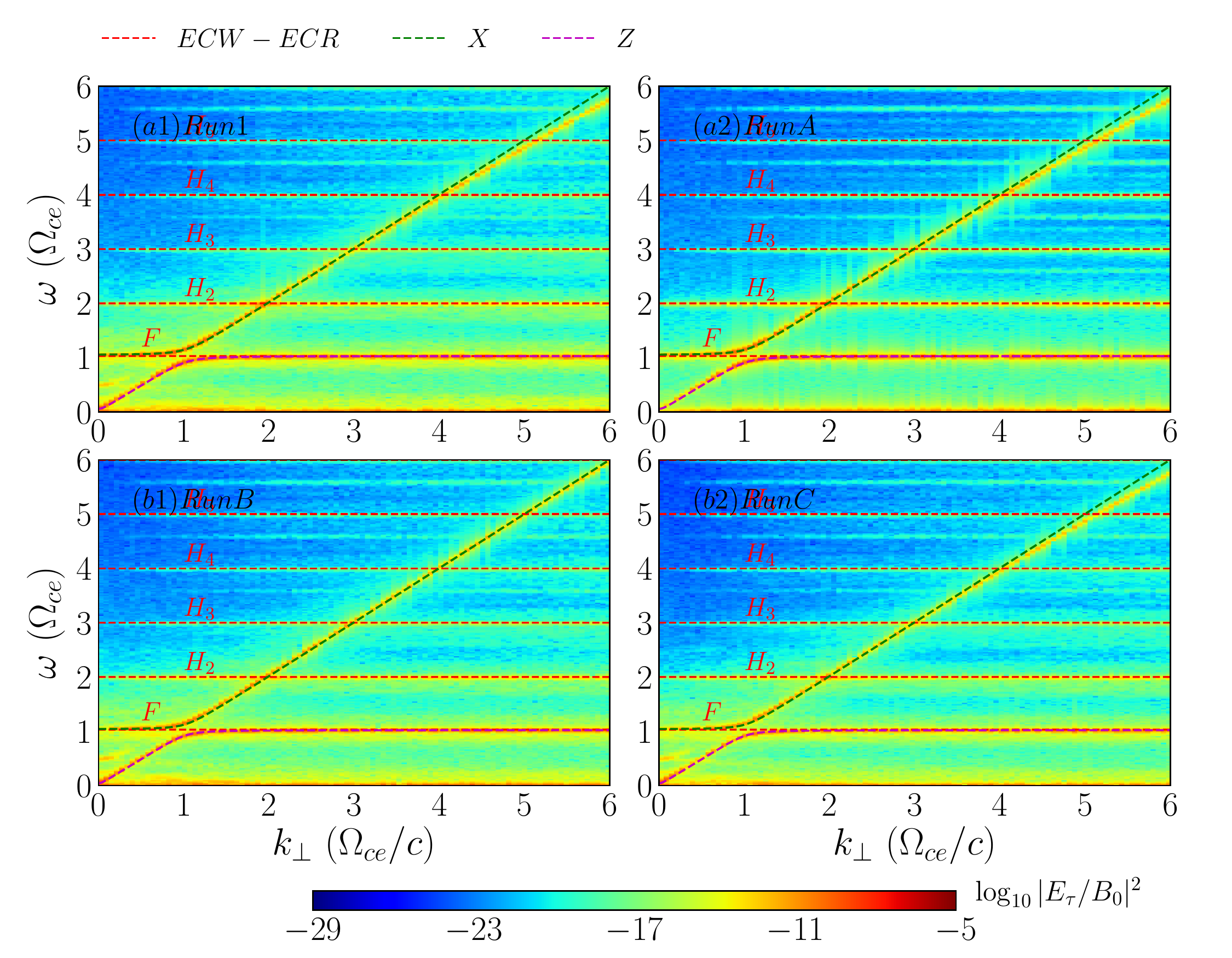}
    \caption[]{PSD in $k_{\perp}\ vs.\ \omega$ plane for Runs 1, A, B and C in the time period $t=100-164\ \omega_{pe}^{-1}$. PSD is evaluated by $\log_{10}|E_{\tau}/B_0|^2$. Dispersion relation curves of fundamental (F) and harmonics (H) of electron cyclotron wave (ECW) modes due to electron cyclotron resonances (ECRs, denoted by red dashed lines), X (green dashed curve) and Z (magenta dashed curve) modes are overlaid.}
    \label{fig:Energy_CT_Et_kperp}
\end{figure}

\myreffig{fig:Energy_CT_El_kpara} shows the power spectral density of the longitudinal electric field in the $(k_{\parallel},\omega)$ domain, in which Langmuir waves are expected to be observed, at the time interval $t=100-164\ \omega_{pe}^{-1}$. For RunA where there is no electron beam (thus with a zero beam-to-background density ratio), only electrostatic Langmuir waves are observed (see \myreffig{fig:Energy_CT_El_kpara}(a2)), while for all other cases where the beam is present, fundamental Langmuir waves and up to the third harmonic of Langmuir waves above the noise level are observed (see \myreffig{fig:Energy_CT_El_kpara}(a1,b1,b2)). We also utilized RunA on the main text of our paper to evaluate the numerical noise level of the power spectra of the different wave modes.

\myreffig{fig:Energy_CT_Et_kperp} shows a comparison of the power spectral density of transverse electric field in the $(k_{\perp},\omega)$ domain, in which electron cyclotron waves caused by ECME are expected to be observed, at the same period. However, only electron cyclotron waves caused by ECRs, X and Z modes are observed for all simulations because there are no perpendicular sources of free energy to cause ECME for the Maxwellian beams. We refer to these wave modes generated in RunA (with only background plasma) as background plasma wave modes, e.g., those wave modes shown in \myreffig{fig:Energy_CT_El_kpara} (a2) and \myreffig{fig:Energy_CT_Et_kperp} (a2).

By comparing results of Run1 and RunA shown in both \myreffig{fig:Energy_CT_El_kpara} and \myreffig{fig:Energy_CT_Et_kperp}, we find that the beam does not significantly distort these background plasma wave modes. We have arguments to think that this distorsion should be small.
The main argument is that the beam is localized in space to a very small region of the simulation domain, so it does not affect the background plasma modes that much in comparison with the more typical case of an non-localized homogeneous beam in the whole simulation domain. The relatively high-beam density indeed contributes towards a larger deviation, but the overall effect is still relatively small. 
The second argument is that our results indicates that any deviation between the normal plasma modes due to the beam should be near the grid resolution, which is  $\Delta \omega=0.047\ \omega_{pe}$ and $\Delta k=0.03\ \omega_{pe}/c$, respectively.
A more accurate estimation for the deviations of the normal plasma modes due to the presence of the beam should be based on a numerical solution of the linearized Vlasov equation via, e.g., a linear Vlasov dispersion solver, at least to provide an upper bound for the case of a homogeneous beam, but this is beyond the scope of our study.

By comparing results of RunB in both \myreffig{fig:Energy_CT_El_kpara} and \myreffig{fig:Energy_CT_Et_kperp} we can now assess the influence of the spatial resolution on wave modes. 
Note that RunB does not only feature a grid cell size half of that of Run1, but also a timestep half of that of Run1 in order to satisfy the CFL condition to the same precision. 
As a result, the RunB is 8 times computationally more expensive per particle than Run1 because it has 4 times the number of cells and twice the number of timesteps.
We notice that in the large wavevector regime the analytical dispersion relation curve of X mode slightly deviates the corresponding PSD enhancement for each simulation (compare the top right section of the panels (a1) and (b1) in \myreffig{fig:Energy_CT_Et_kperp}). As mentioned in the main text (see Section~\ref{sec:electroncyclotronwaves}), this numerical effect has to do with the fact that any numerically obtained dispersion relation curve differs from the corresponding analytical (continuous) dispersion relation curve due to the use of a finite grid cell size and finite time step. Because a smaller grid cell size and time step are used in RunB, the analytical dispersion relation curve of the X mode fits the numerically obtained X-mode curve better than in the other simulations with lower resolution. We find that there is not a significant visual difference in the power spectral density of the longitudinal electric field in the $(k_{\parallel},\omega)$ domain and that of the transverse electric field in the $(k_{\perp},\omega)$ domain between RunB and Run1. 

Similarly, by comparing results of RunC in \myreffig{fig:Energy_CT_El_kpara} and \myreffig{fig:Energy_CT_Et_kperp} we see that the numerically obtained dispersion curves barely change by using the double number of macro-particles per cell.

\begin{figure}
    \centering
    \includegraphics[width=0.85\textwidth]{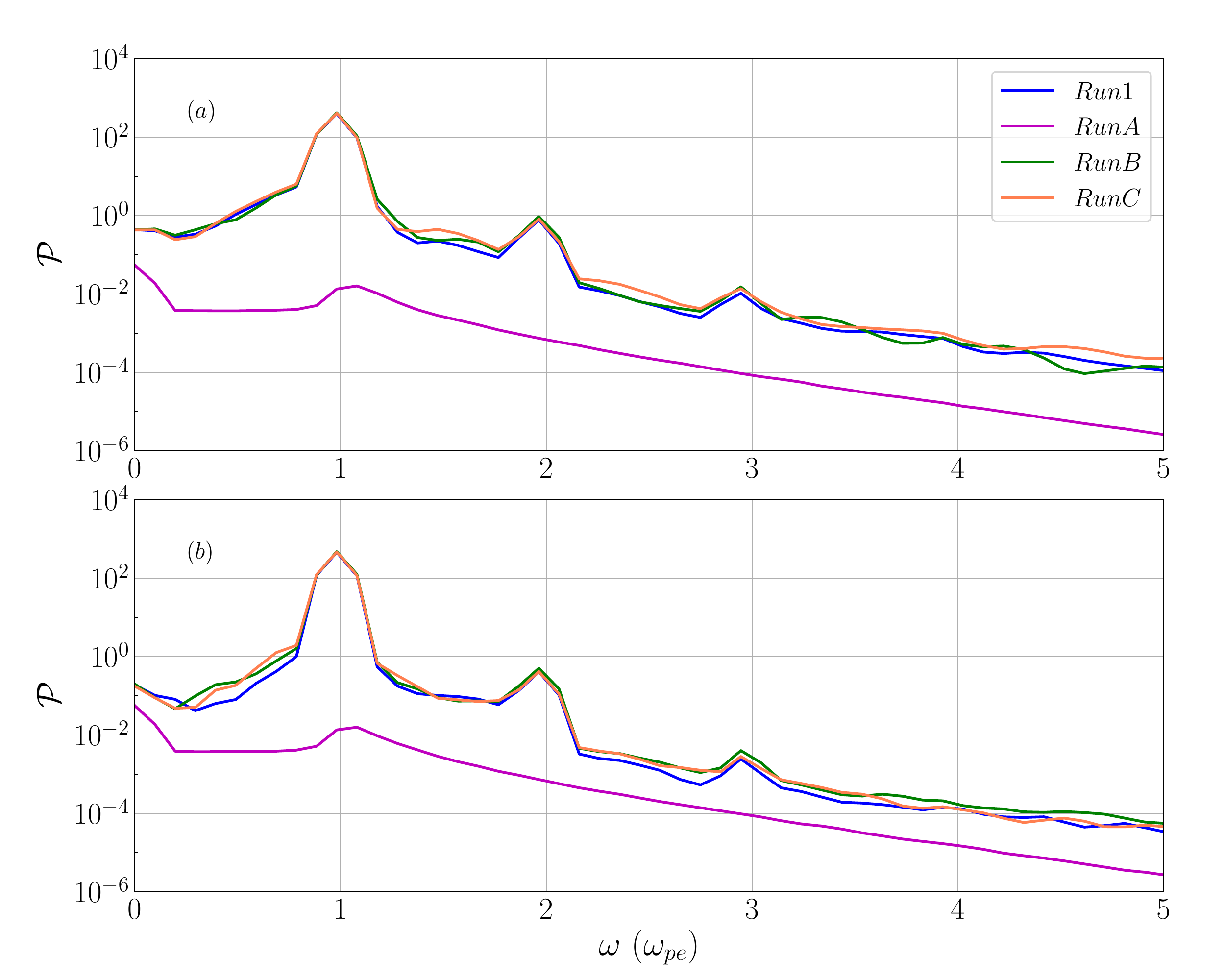}
    \caption[]{Power spectra $\mathcal{P}(\omega)$ corresponding to \myreffig{fig:Energy_CT_El_kpara} of the convergence simulation runs RunA, B, C and the standard Run1 during the time periods (a) $t=50-114\ \omega_{pe}^{-1}$ and (b) $t=100-164\ \omega_{pe}^{-1}$ respectively.}
    \label{fig:Pw_CT_linear}
\end{figure}

The previous comparison was rather visual and qualitative. 
In order to address to what extent the wave modes are actually modified between the various convergence runs, we quantitatively measured this difference by using the integrated spectral power for simulations. 
\myreffig{fig:Pw_CT_linear} shows the integrated spectral power $\mathcal{P}(\omega)$ according to \myrefeq{eq:EnergyLongitudinalEl} of the parallel electric field corresponding to \myreffig{fig:Energy_CT_El_kpara} at the time periods $t=50-114\ \omega_{pe}^{-1}$ and $t=100-164\ \omega_{pe}^{-1}$, respectively. The local maxima of those curves indicate the characteristic frequencies at the maximum spectral power of beam-generated Langmuir waves and their harmonics (i.e., $\omega=\omega_{pe},2\omega_{pe},3\omega_{pe}$). 
In general, \myreffig{fig:Pw_CT_linear} shows that the integrated spectral power curves of Run1, RunB and RunC almost overlap with each other. The curve corresponding to RunA case (see magenta curve in \myreffig{fig:Pw_CT_linear}(a,b)), where there is no electron beam, is significantly lower than the other curves with beam.

We proceeded to quantify the relative difference between those curves by calculating $\Delta \mathcal{E}_i=(\mathcal{E}_i-\mathcal{E}_1)/\mathcal{E}_1$, where $\mathcal{E}_i$ is the total power density defined as $\mathcal{E}_i=\displaystyle\int\mathcal{P}_i(\omega)d\omega,\ (i=1,B,C)$. 
For the time period $t=50-114\ \omega_{pe}^{-1}$, the relative difference of the cases B ($\Delta \mathcal{E}_B$) and C ($\Delta \mathcal{E}_C$) with respect to the standard Run1 amounts to  $5.6\%$ and $3.2\%$, respectively. While for the time period $t=100-164\ \omega_{pe}^{-1}$ this difference is equal to  $4.8\%$ and $3.1\%$ for runs B and C, respectively.

In order to compare how different  the spectral power at specific frequencies is, we calculated the relative difference $\Delta \mathcal{P}_i(\omega_{pe})=\left(\mathcal{P}_i(\omega_{pe})-\mathcal{P}_1(\omega_{pe})\right)/\mathcal{P}_1 ,\ (i=B,C)$. Note that this calculation is performed for the first spectral power peak $\omega_{pe}$ corresponding to the fundamental Langmuir mode.
This calculation reveals that for the time period $t=50-114\ \omega_{pe}^{-1}$, the relative difference of the case B and C with respect to the standard Run1 amounts to  $5.1\%$ and $2.0\%$, separately. While for the time period $t=100-164\ \omega_{pe}^{-1}$ this difference is equal to $4.4\%$ and $2.3\%$ for runs B and C, separately.

We can thus conclude that the spectral power varies by a maximum factor of $5-6\%$ by improving either the grid resolution or the number of particles per cell. This means that there are variations in the waves generated in simulations with improved resolution, but they are rather small. In general, the original values of the spatial and time resolution, as well as the number of macro-particles per cell (as these in Run1) are sufficient to resolve the fundamental and harmonics of waves to the level required for our study.

\bibliographystyle{jpp}
\bibliography{./library_manu.bib}

\end{document}